\documentclass[aps,pre,twocolumn,groupedaddress,amsmath]{revtex4-1}
\usepackage{color}
\usepackage{graphicx}% Include figure files
\usepackage{dcolumn}% Align table columns on decimal point
\usepackage{bm}% bold math

\begin{document}

%\preprint{}

\title{Controlled nanochannel lattice formation utilizing prepatterned substrates}


\author{Michael H. K\"opf}\email{m.koepf@uni-muenster.de}
\author{Svetlana V. Gurevich}
\author{Rudolf Friedrich}
\affiliation{%
Institute for Theoretical Physics, University of M\"unster, Wilhelm-Klemm-Str.~9, D-48149 M\"unster, Germany\\
}

\date{\today}

\begin{abstract}
Solid substrates can be endued with self-organized regular stripe patterns of nanoscopic lengthscale by Langmuir-Blodgett transfer of organic monolayers. Here we consider the effect of periodically prepatterned substrates on this process of pattern formation. It leads to a time periodic forcing of the oscillatory behavior at the meniscus. Utilizing higher order synchronization with this forcing, complex periodic patterns of predefined wavelength can be created. The dependence of the synchronization on the amplitude and the wavelength of the wetting contrast is investigated in one and two spatial dimensions and the resulting patterns are discussed. Furthermore, the effect of prepatterned substrates on the pattern selection process is investigated.
\end{abstract}

%\keywords{}

\maketitle

\section{Introduction}
Prepatterned substrates are widely used as means to control wetting and dewetting processes~\cite{GHLL_Science_99,QXY_AdvMat_99,SOG_AdvMat_02,CGO_PRL_06}. Chemically or topographically prestructured, the substrate exhibits heterogeneous wetting properties and thereby enforces an extrinsic pattern upon a covering liquid layer. A particularly interesting scenario emerges if the liquid itself exhibits pattern formation \emph{intrinsically}, yielding regular or irregular structures even on homogeneous substrates. This is the case, for example, in Langmuir-Blodgett transfer of organic and metallic monolayers~\cite{GCF_Nature_00,HJ_NatMater_05} and in controlled evaporation of nanoparticle suspensions~\cite{YS_AdvFuncMat_05}. On a prepatterned substrate, this gives rise to the formation of a final pattern resulting from a competition between the extrinsic structure imposed by the substrate and the intrinsic structure that is formed naturally in the system.

Here, we want to focus on the effect of \emph{periodically} prepatterned substrates on the transfer of monolayers of amphiphilic molecules like pulmonary surfactant DPPC (dipalmitoylphosphatidylcholine). Using homogeneous substrates, this transfer process results in the spontaneous formation of nanochannel lattices, that is, regular stripe patterns with a typical wavelength of merely a few hundred  nanometers~\cite{GCF_Nature_00,CLH_AccChemRes_07,KCR_EPS_94}. The corresponding experiments utilize the Langmuir-Blodgett method, where the substrate is withdrawn from a trough filled with water on which a monolayer has been prepared, and are conducted under conditions close to the first order phase transition between the liquid-expanded (LE) and the liquid-condensed (LC) phases of the monolayer. The observed patterns consist of ordered arrays of LE and LC domains, arranged into stripes oriented either parallel or perpendicular to the contact line. For certain experimental parameters, one can even obtain a combination of both orientations, resulting in a rectangular structure.
The formation of the nanochannels results from a substrate-monolayer interaction, commonly referred to as substrate-mediated condensation (SMC)~\cite{LBM_CollSurfA_00,GR_CollSurfA_98,SR_Langmuir_94,KCR_EPS_94,RS_ThinSolidFilms_92}, and can be understood theoretically as a phase decomposition process under influence of the corresponding interaction field~\cite{KGFC_Langmuir_10}.

Since the menisucs, where the condensated domains are generated, moves across the prepatterned substrate, the spatially periodic prestructure is translated into a spatiotemporal forcing. The effect of spatiotemporal forcing on different pattern forming systems, e.g., photosensitive Turing systems \cite{RMM_PRL_03}, has been investigated earlier. However, in these cases, an external frequency is imposed in an otherwise nonoscillatory system, whereas even the unforced pattern formation in monolayer transfer is an oscillatory process.

The theoretical investigation of dewetting systems with surfactants is typically based on the lubrication approximation which greatly facilitates the description of thin film flow~\cite{BEI_RevModPhys_09,CM_RevModPhys_09,ODB_RevModPhys_97}. In this framework, surfactant-covered thin liquid films can be modeled by two coupled nonlinear differential equations describing the height profile of the film and the surfactant density~\cite{CM_RevModPhys_09,ODB_RevModPhys_97,MT_PhysFluids_97,WCM_PhysFluids_02}, whereas prepatterned substrates can be described by a spatially varying disjoining potential~\cite{TK_PRL_06,TBBB_EPJE_03}. In addition, monolayers in the vicinity of the LE-LC phase transition can be modeled by coupling of the lateral pressure, that is the surface tension, to a Cahn-Hilliard type free-energy functional~\cite{KGF_EPL_09}. The substrate-mediated condensation can then be described by an interaction field acting on the monolayer~\cite{KGFC_Langmuir_10}, whose strength depends on the distance to the substrate, that is, the thickness of the liquid film dividing substrate and monolayer. Due to this interaction, the condensed state is energetically favored when the monolayer is close to the substrate, leading to the partial condensation necessary to form the alternating LE/LC stripes.

Investigating the influence of periodically prepatterned substrates on this pattern formation process %in one and two spatial dimension 
we shall find synchronization phenomena~\cite{PRKsync} of various orders, leading to the formation of complex periodic patterns depending on the amplitude and the wavelength of the substrate's wetting contrast. We will further show that a periodic prepattern can be used to inhibit an orientation transition from stripes arranged parallel to the contact line to perpendicular stripes and even allows to produce new structures such as well-ordered domains of circular LE domains alternating with LE/LC stripes.
\section{Governing equations}
\subsection{Thin film flow}
\begin{figure}
\includegraphics[width=.48\textwidth]{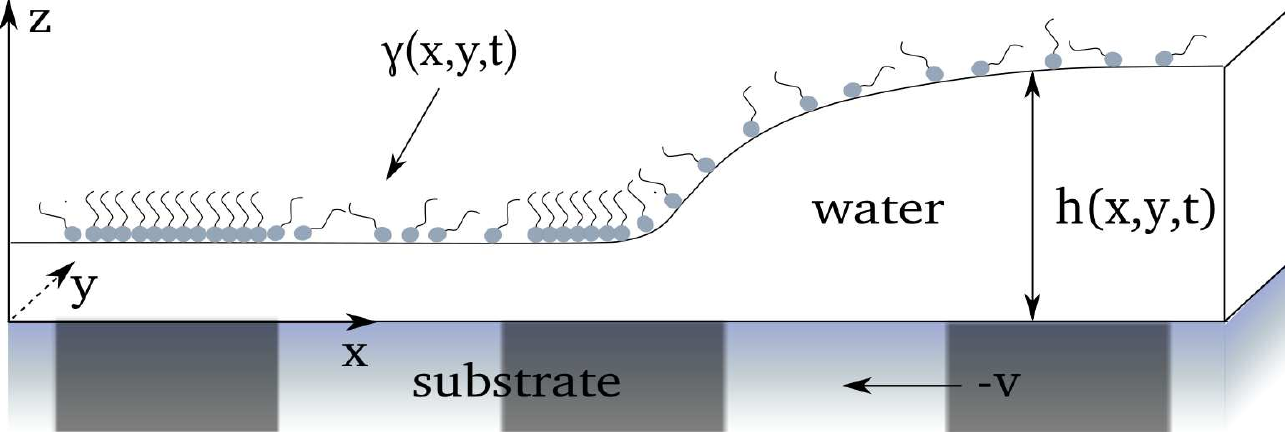}
\caption{\label{fig:drawing_prep}Schematic drawing of a surfactant-covered meniscus on a prepatterned substrate. The height profile $h(x,y,t)$ indicates the film thickness at location $(x,y)$ and time $t$, whereas $\gamma(x,y,t)$ describes the surfactant density at the surface above $(x,y)$.}
\end{figure}
Let us consider an evaporating thin film of water on a periodically prestructured substrate which is withdrawn with constant velocity $v$ in negative $x$-direction from a water reservoir which ensures a constant film height at the right boundary of the system (see figure \ref{fig:drawing_prep}). A meniscus is then given by the balance of evaporation and the supply of fresh water, which is carried from the reservoir by the moving plate, yielding a contact line moving relatively to the substrate. The water reservoir is assumed to be covered with a surfactant monolayer of constant density, which is then carried by the flow towards the contact line. 

The liquid film is described by its height profile $h(\bm{x},t)$, which indicates the local film thickness, whereas the surfactant density at the surface above the point $\bm{x}=(x,y)$ is described by the function $\gamma(\bm{x},t)$. 
Introducing the scales $h_0$, $l_0$, $t_0$ and $\gamma_0$ for height, length, time, and surfactant density, the time evolution equations for the height profile $H(\bm{X},T):=h(\bm{x}/l_0,t/t_0)/h_0$ and the surfactant density $\Gamma(\bm{X},T):=\gamma(\bm{x}/l_0,t/t_0)/\gamma_0$ can be written in dimensionless form as \cite{KGFC_Langmuir_10}
\begin{align}
\partial_T H &= -\nabla\cdot \left\{\frac{H^3}{3}\nabla\bar{P} + \frac{H^2}{2}\nabla \hat{\Sigma} - V\bm{e}_X H\right\} 
-\mathrm{Ev}\,\delta\mu,\quad \label{evolh} \\
\partial_T \Gamma &= -\nabla\cdot\left\{ \frac{\Gamma H^2}{2}\nabla\bar{P} +\Gamma H \nabla \hat{\Sigma} - V\bm{e}_X\Gamma\right\}, \label{evolgamma}
\end{align}
with the generalized pressure $\bar{P}=\epsilon^3\mathrm{Ca}^{-1}\hat{\sigma}\nabla^2H-\Pi(H)$ 
and $\hat{\Sigma}=\epsilon\mathrm{Ca}^{-1}\hat{\sigma}$. Here, the parameter $\epsilon=h_0/l_0$ signifies the ratio of the characteristic height- and length scales of the system and is assumed to be small in thin film problems. The inverse capillary number $\mathrm{Ca}^{-1}=\sigma_0/(\eta U_0)$ is the ratio of a characteristic surface tension $\sigma_0$, the dynamic viscosity $\eta$ and the characteristic velocity $U_0=l_0/t_0$. The scaled surface tension $\hat{\sigma}=\sigma/\sigma_0$ of the water film depends on the density $\Gamma$ of the surfactant monolayer. Therefore, spatial variations of $\Gamma$ lead to spatial variations of $\hat{\sigma}$, inducing Marangoni forces which are accounted for by the terms $\sim\nabla\hat{\Sigma}$ in the equations \eqref{evolh} and \eqref{evolgamma}. The dependence of $\hat{\sigma}$ on the monolayer density can be described in terms of a lateral pressure~\cite{Adamson} and, as will be outlined in the next section, is deeply related to the surfactant thermodynamics.

Since the substrate is withdrawn with a constant velocity $v$, both equations \eqref{evolh} and \eqref{evolgamma} include an advective term involving $V=v/U_0$.

Note, that the generalized pressure $\bar{P}$ comprises not only the Laplace pressure $\sim\nabla^2 H$, which is present in every free fluid interface problem, but also the disjoining pressure $\Pi(H)$ due to the interaction of substrate and liquid. These interactions become significant for liquid films of heights lower than a few hundred nanometers. In the literature, different expressions for $\Pi(H)$ have been considered (see \cite{ODB_RevModPhys_97} and references therein for a discussion of possible choices). Here, we employ the dimensionless expression
\begin{equation}
\Pi(H,\bm{X}) = M(\bm{X})\left(A_3 H^{-3}- A_6 H^{-6}\right) \label{disjoining}
\end{equation}
where $A_3=\epsilon^2 t_0 a_3/ (h_0^3\eta),A_6=\epsilon^2 t_0 a_6/ (h_0^6\eta)$ with Hamaker constants $a_3,a_6>0$, yielding, for $M=\mathrm{const.}$, a precursor height defined by $\Pi(H_\mathrm{p})=0\Leftrightarrow H_p=\sqrt[3]{A_6/A_3}$. The function $M(\bm{X})$ accounts for the spatially heterogeneous wettability of the prepatterned substrate. This approach is similar to the one in \cite{TBBB_EPJE_03,TK_PRL_06}. Since we are interested in periodically prestructured substrates, we choose $M(\bm{X})$ to be a periodic modulation
\begin{equation*}
M(\bm{X})=1+\rho\,\xi\left( X+VT \right) 
\end{equation*} 
which moves along with the substrate and has the form of a kink-antikink train
\begin{equation*}
\xi(X)=\tanh\left(\frac{10\,\mathrm{tri}(X/L_\mathrm{p})}{\kappa_\mathrm{p}}\right)
\end{equation*}
with the periodic triangle function
\begin{equation}
\mathrm{tri}(x)=1-2\left|\{x\}-\frac{1}{2}\right|,
\end{equation}
where $\{x\}:=x \mod 1$ denotes the fractional part of $x$.
This means, that we have a modulation with amplitude $\rho$ and wavelength $L_\mathrm{p}$. 
For $\rho\rightarrow 0$ the homogeneous substrate is obtained. 
The constant $\kappa_\mathrm{p}$ determines the sharpness of the contrast. It has to be noted, that very small $\kappa_\mathrm{p}$ would violate the assumption of slow variation in $x$- and $y$-direction, which is the basis of the lubrication approximation.

Due to evaporation, a sink term is present in the evolution equation for $H$, so that the fluid volume is not conserved. Here $\delta\mu=\mu_\mathrm{w}-\mu_\mathrm{v}$ denotes the difference of the chemical potentials of the water film and the ambient vapor phase, whereas $\mathrm{Ev}=\eta l_0^2 Q_\mathrm{e}/ h_0^3$ is the evaporation number with effective rate constant $Q_\mathrm{e}$. The pressure in the vapor above the film is assumed to be close to the saturation pressure, allowing us to identify the chemical potential of the water film with the negative generalized pressure~\cite{Pi_PRE_04,TVA_JPhysCM_09}, that is $\mu_\mathrm{w}=-\bar{P},\mu_\mathrm{v}=\mathrm{const}$.

It has to be noted that monolayer rigidity is not taken into account in the model given by eqs. \eqref{evolh} and \eqref{evolgamma}. This is due to the fact, that, going from low to high densities, a significant surface rigidity $K_\mathrm{c}$ is observed experimentally only after the LC to solid (S) phase transition. 
To give a more precise estimation of the influence of surface rigidity,
one can supplement lubrication theory by a contribution resulting from minimization of the corresponding Helfrich energy of a stiff interface. As a result, an addtional quantity $\sigma_2=K_\mathrm{c}/l_0^2$ enters the equations with exactly the same prefactors as the surface tension $\sigma$. Thus, $\sigma_2$ can be regarded as a second surface tension and can be directly compared in its magnitude to $\sigma$.
Even in the S phase, which is significantly more rigid than the liquid-condensed phase considered in our problem, the bending rigidity of a DPPC monolayer is only $K_\mathrm{c}\le 70\,k_\mathrm{B} T$ \cite{DBB_JPhysCondMat_90}, that is of order $10^{-19}\,\mathrm{J}$. Since $l_0\approx 10^{-7}\,\mathrm{m}$ one can estimate that in this case $\sigma_2\approx 10^{-5}\, \mathrm{J}/\mathrm{m}^2$, which is three orders of magnitude below the surface tension of water and therefore neglectable. For other materials, however, the rigidity can be
significantly larger, so that it might have to be taken into account for an
accurate description of thin films covered with such monolayers. 
That is, the applicability of the description presented here is  limited to materials where the rigidity $K_\mathrm{c}$ of the monolayer is less than a few hundred $k_\mathrm{B} T$.
\subsection{Surfactant thermodynamics and substrate mediated condensation}
The flow of a thin liquid film on a substrate depends on the surface tension $\sigma$ at the free liquid-vapor interface. In the presence of a surfactant monolayer, the surface tension varies across the surface, depending on the local surfactant density $\Gamma$ and its spatial derivatives. This dependence is described by the lateral pressure $P_\mathrm{lat}$ and the relation ~\cite{Adamson}
\begin{equation}
\hat{\sigma} = \hat{\sigma}_\mathrm{abs} - P_\mathrm{lat},  \label{platdef}
\end{equation}
where $\hat{\sigma}_\mathrm{abs}=\sigma_\mathrm{abs}/\sigma_0$ denotes the 
scaled surface tension in absence of any surfactant.

The relation between lateral pressure and surfactant density is determined by an equation of state which can be obtained from the free-energy of the monolayer. Since the density $\Gamma$ may vary across the surface, its free-energy can be written in the form proposed by Cahn and Hilliard for a general non-homogeneous system~\cite{CH_JChemPhys_58}, i.e.~ $\mathcal{F}[\Gamma]=\int \mathrm{d}^2X F(\Gamma(\bm{X}))$ with a free-energy density
\begin{equation}
F=\epsilon^2\frac{K}{2}(\nabla\Gamma)^2 
+ F_\mathrm{hom}(\Gamma-\Gamma_\mathrm{cr}).
\end{equation}
Here, $K=\kappa\gamma_0^2/(h_0^2\sigma_0)$ denotes the nondimensionalized form of the line tension $\kappa$ between domains of different densities, while the function $F_\mathrm{hom}$ represents the free-energy of a system with homogeneous surfactant density $\tilde{\Gamma}$. To focus on the LE-LC first-order phase transition of the monolayer, we consider a bistable system where the low- and high-density minima of $F_\mathrm{hom}$ correspond to the liquid-expanded and the liquid-condensed phase, respectively. For the sake of simplicity we assume the function $F_\mathrm{hom}$ to be a symmetric double-well, centered around a critical density $\Gamma_\mathrm{cr}$. It has to be emphasized, that this assumption is no real limitation, since, around the coexistence plateau, a wide range of experimentally obtained pressure-area isotherms can be fitted under the assumption of a symmetric $F_\mathrm{hom}$. Using $\tilde{\Gamma}=\Gamma-\Gamma_\mathrm{cr}$, we write
\begin{equation}
F_\mathrm{hom}(\tilde{\Gamma})=F_\mathrm{cr} + \frac{M_1}{4M_2^2}\tilde{\Gamma}^2\left(\tilde{\Gamma}^2-2M_2^2\right). \label{fhom}
\end{equation}
Here, $M_1$ equals the curvature of the function $F_\mathrm{hom}$ in its center $\tilde{\Gamma}=0$ and $M_2$ denotes the difference between the densities $\Gamma_\mathrm{LE}$,$\Gamma_\mathrm{LC}$ of the pure LE and LC phases and the critical density $\Gamma_\mathrm{cr}$, that is $M_2:=|\Gamma_\mathrm{LE}-\Gamma_\mathrm{cr}|=|\Gamma_\mathrm{LC}-\Gamma_\mathrm{cr}|$. The constant $F_\mathrm{cr}$ is the free-energy of a 
homogeneous monolayer with the critical density $\Gamma_\mathrm{cr}$.

The lateral pressure can be obtained from the surfactant equation of state, which is related to the free-energy by
\begin{equation}
P_\mathrm{lat} = -F + \Gamma\mu,\label{p_lat_dimensionful}% = -f+\gamma\frac{\delta\mathcal{F}}{\delta\gamma}.
\end{equation}
where the chemical potential $\mu$ is obtained from $\mathcal{F}$ by functional derivation with respect to $\Gamma$, yielding
\begin{equation}
\mu=\frac{\delta\mathcal{F}}{\delta\Gamma}=-\epsilon^2K \nabla^2\Gamma + \frac{\partial F_\mathrm{hom}}{\partial \Gamma}.
\end{equation}
Inserting eqs. \eqref{fhom} and \eqref{p_lat_dimensionful} into \eqref{platdef}, 
the physical meaning of $F_\mathrm{cr}$ becomes obvious:
\begin{equation*}
F_\mathrm{cr} = \hat{\sigma}(\Gamma_\mathrm{cr})-\hat{\sigma}_\mathrm{abs}.
\end{equation*}
%
%Only in homogeneous systems the lateral pressure is completely described by equation \eqref{p_lat_dimensionful}. 
The above description is still incomplete as inhomogeneous systems are considered.
In case of a spatially varying density $\Gamma$ 
the pressure is given by a tensor $\bm{P}$ with the components~\cite{DS_AdvChemPhys_82,Ev_AdvPhysics_79}
\begin{equation}
P_{ij}=P_\mathrm{lat}\delta_{ij}+\epsilon^2 K(\partial_i\Gamma)(\partial_j\Gamma).\label{p_tens}
\end{equation}
However, in the Laplace pressure, which is of order $\epsilon^3$, the derivative terms of $\bm{P}$ are of fifth order in $\epsilon$. Since $\epsilon$ is assumed to be small, they can be safely ignored. In the Marangoni term $\nabla\hat{\Sigma}$, the lateral pressure enters in the form $\nabla \cdot\bm{P} = \nabla \left(P_\mathrm{lat}+\epsilon^2K(\nabla\Gamma)^2 \right)$.
Therefore, one can simply use the \emph{scalar} pressure 
\begin{equation}
P = P_\mathrm{lat}+\epsilon^2K(\nabla\Gamma)^2 
\end{equation}
throughout all calculations.

It is an experimentally well-established fact that interaction with the substrate influences the thermodynamic behavior of the monolayer, when the liquid subphase dividing them becomes very thin~\cite{LBM_CollSurfA_00,GR_CollSurfA_98,SR_Langmuir_94,KCR_EPS_94,RS_ThinSolidFilms_92}. It is by this influence, that a pure monolayer condensates at the substrate even under conditions where it would still be perfectly stable on a subphase of $\sim 100\,\mathrm{nm}$ thickness. This substrate-mediated condensation effect can be modeled by inclusion of an external field in the free-energy of the monolayer, whose strength depends on the distance to the substrate which, in our case, is given by the thickness of the water film $H$. The free-energy density in presence of the substrate-monolayer interaction is then given by
\begin{equation}
F^\mathrm{(SMC)}_\mathrm{hom}=F(\tilde{\Gamma}) + S(H)\tilde{\Gamma}. \label{F_hom_SMC}
\end{equation}
Here, $S(H)$ describes the functional dependence of the strength of the substrate-mediated condensation on the distance. Unfortunately, the exact form of $S(H)$ still remains to be determined by experiment or from a microscopic theory. Nevertheless, there exist rough estimates of the overall lateral pressure difference caused by SMC: For DPPC and DMPE the LE-LC coexistence pressure at the substrate can be several millinewton per meter lower than for a floating monolayer~\cite{SR_Langmuir_94}, yielding an overall difference of more than $40\%$. In terms of $S(H)$ this means that $S(H)$ goes to a finite negative value for $H\rightarrow H_\mathrm{p}$, leading to a tilt of $F_\mathrm{hom}$ towards its higher density  minimum, that is the LC phase, as the monolayer approaches the substrate. Also, $S(H)$ should vanish quickly for $H\rightarrow\infty$, since the SMC can only be observed very close to the substrate. 

Although the results presented in this article are qualitatively valid for a class of functions $S(H)$ meeting these minimal constraints, without loss of generality we choose $S(H)=B \Psi(H)$, where $\Psi(H)=\int d H'\Pi(H')$ with integration constant zero is the potential of the substrate-liquid interaction and $B$ is a positive coupling constant. By this choice, SMC acts on length scales comparable to the substrate-liquid interaction.
\subsection{Choice of scales}
So far, we did not specify a certain set of scales $h_0$, $l_0$, $t_0$, and $\gamma_0$. A convenient choice results from demanding $\epsilon^3\mathrm{Ca}^{-1}=\epsilon^2 K = H_\mathrm{p}=1$. This means, that we use the precursor height as a natural height scale of the system by setting $h_0=h_\mathrm{p}$. Due to this choice we have $A:=A_3=A_6=(\gamma_0^2 \kappa a_3)/(\sigma_0^2 h_\mathrm{p}^4)$ in equation \eqref{disjoining}.

The characteristic length scale $l_0$ of the system is then given by the ratio of the surface tension $\sigma_0$ and the domain line tension $\kappa$ as $l_0=\gamma_0 \sqrt{\kappa/\sigma_0}$ and the natural time scale of the problem is obtained from the relation $t_0=\gamma_0^4\kappa^2\eta / (h_0^3 \sigma_0^3)$.
With this choice of scales, $\epsilon$ obtains a direct physical meaning as
\begin{equation}
\epsilon=\sqrt{\frac{\sigma_0}{\gamma_0^2\kappa}\left(\frac{a_6}{a_3} \right)^{2/3}} %\gamma_0\sqrt{\frac{\kappa}{\sigma_0}}\sqrt[3]{\frac{a_3}{a_6}}. 
\end{equation}
relates the forces acting at the monolayer covered surface, characterized by $\sigma_0,\gamma_0$, and $\kappa$, to the forces acting between substrate and liquid, determined by the constants $a_3$ and $a_6$.

Since we are interested in systems in vicinity of the LE-LC phase transition, it is convenient to set the scale $\gamma_0$ equal to the critical surfactant density, that is $\gamma_0=\gamma_\mathrm{cr}$ and to define the characteristic surface tension  $\sigma_0=\sigma(\gamma_\mathrm{cr})$ as the surface tension of water covered with a monolayer of density $\gamma_\mathrm{cr}$. 
The resulting scaled expression for the surface tension is given by
\begin{equation}
\hat{\sigma}=1-\epsilon^2 P_\mathrm{hom}+\left[\Gamma\Delta\Gamma-\frac{1}{2}\left(\nabla\Gamma\right)^2 \right]
\end{equation}
with
\begin{align}
P_\mathrm{hom} &=F_\mathrm{cr}-F_\mathrm{hom}+\Gamma\frac{\partial F_\mathrm{hom}}{\partial\Gamma} \\
&=\frac{M_1}{M_2^2}\tilde{\Gamma}\left(\tilde{\Gamma}^2-M_2^2\left(1+\frac{\tilde{\Gamma}}{2}\right) \right) + S(H). \label{P_hom}
\end{align}
Analogously, the generalized pressure $\bar{P}$ in the equations \eqref{evolh} and \eqref{evolgamma} is obtained as
\begin{equation}
\bar{P}=(1-\epsilon^2 P_\mathrm{hom})\nabla^2 H - \Pi(H) 
\end{equation}
whereas $\nabla\hat{\Sigma}$ can be written in the form
\begin{equation}
\nabla\hat{\Sigma}=-\nabla P_\mathrm{hom}+\epsilon^{-2}\Gamma\nabla^3\Gamma.
\end{equation}
Note, that $\bar{P}$ does not contain any derivatives of $\Gamma$, due to the argument following equation \eqref{p_tens}.
\section{Transfer onto periodically prepatterned substrates}
\subsection{Methods}
The system \eqref{evolh}-\eqref{evolgamma} is simulated numerically in one and two spatial dimensions using second-order finite differences for discretization of space and an embedded adaptive Runge-Kutta scheme of order 4 (5) for time-stepping~\cite{NRC}. 
In the 1D (2D) case, the integration domain is $[0,L]$ ($[0,L]\times[0,L]$) with $L=1200$, discretized on a grid of $384$ ($384\times 384$) gridpoints. For a more detailed investigation of some of the obtained patterns, a larger domain with $L=2000$ is simulated using $640$ gridpoints. In $X$-direction the boundary conditions are
\begin{equation*}
\Gamma(L)=\Gamma_L, \quad \partial^2_X \Gamma(L)=0 = \partial_X \Gamma(0)=\partial^2_{X} \Gamma(0),
\end{equation*}
\begin{equation*}
H(L)=H_L, \quad \partial^2_X H(L)=0 = \partial_X H(0)=\partial^2_X H(0),
\end{equation*}
with $\Gamma_L=0.835$, corresponding to transfer from a reservoir which is covered with a homogeneous monolayer in the pure LE phase. 
The choice of boundary conditions at $X=0$ is not obvious, since ideally the material would simply leave the integration domain to the left. However, such nonreflective boundary conditions are extremely complicated to realize for equations like \eqref{evolh} and \eqref{evolgamma}. Therefore, we use 
the above simple set of conditions which we tested on three different domanins
with $L=1200,2400,$ and $3600$ without finding any influence of the boundaries on the pattern formation at the meniscus. In the two-dimensional simulations, periodic boundary conditions hold in the $Y$-direction.

We investigate periodically prepatterned substrates with wavelengths $L_\mathrm{p}=200$, $300$, and $400$, each for wetting contrast amplitudes between $\rho=0.001$ and $\rho=0.01$. All three prepattern wavelengths are on the same order of magnitude as the natural wavelengths of the system. For comparison, we will start with a discussion of the case of homogeneous substrates, $\rho=0$.

\begin{figure}
\includegraphics[width=.4\textwidth]{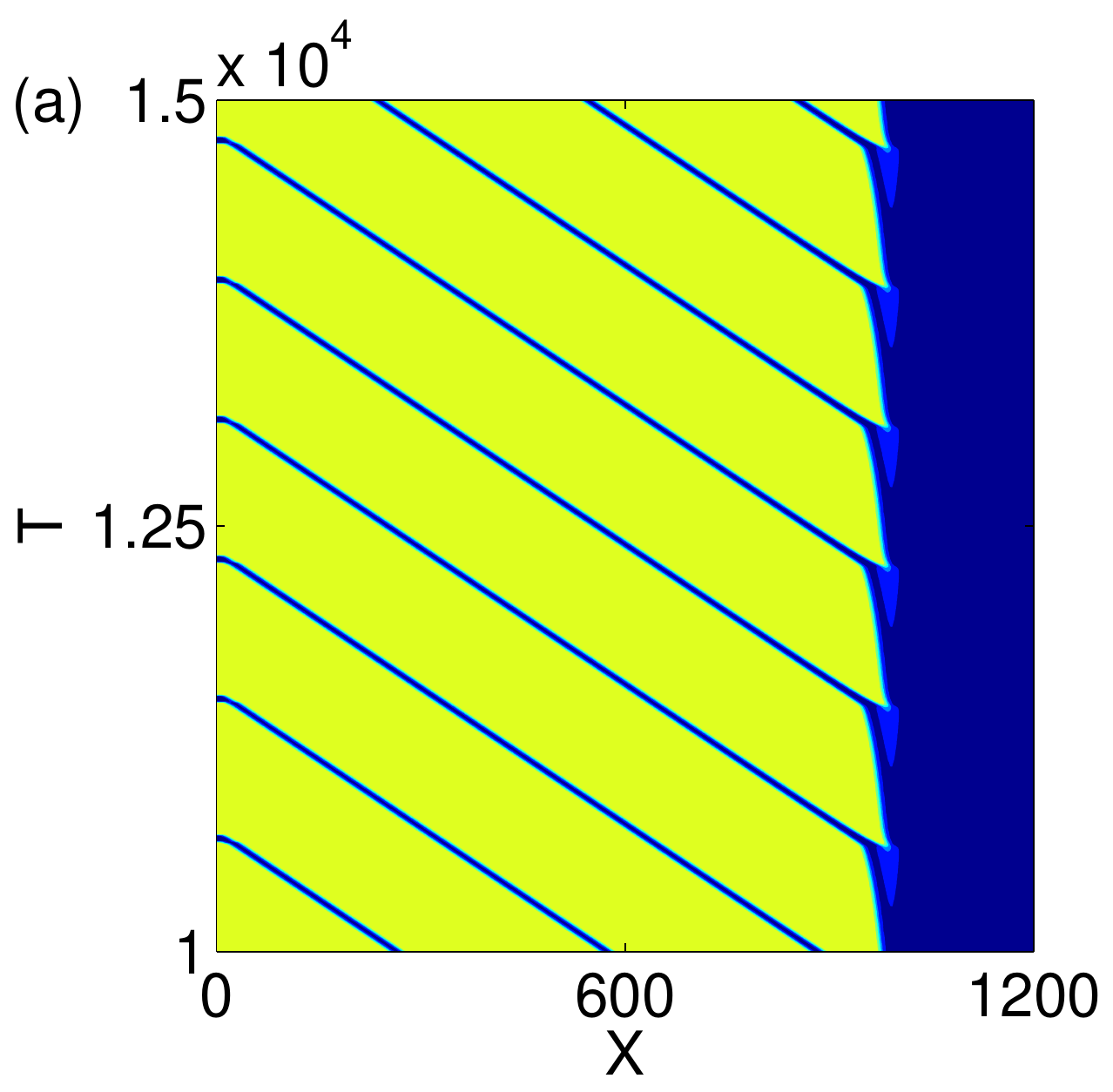}
\includegraphics[width=.4\textwidth]{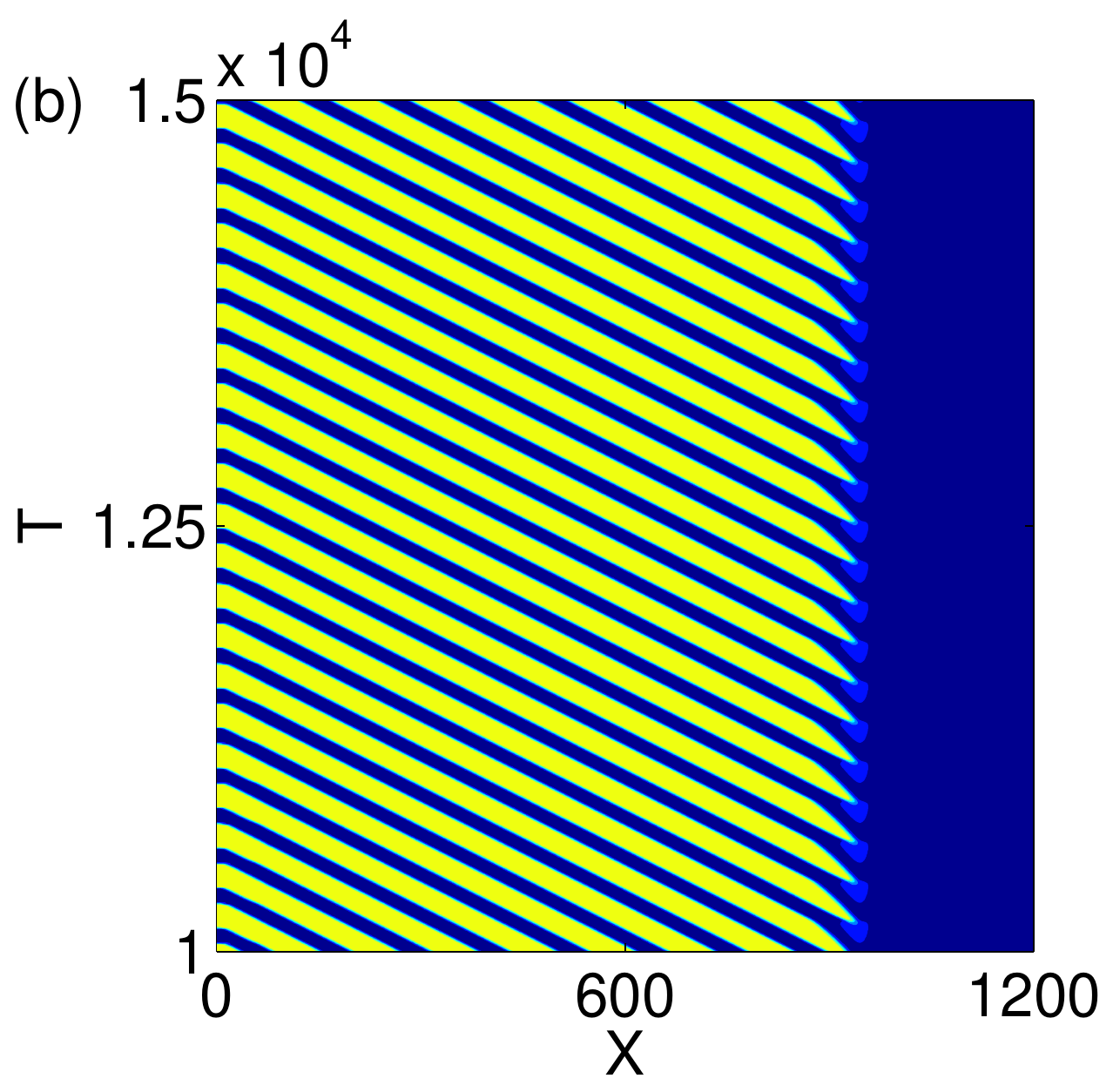}
\includegraphics[width=.4\textwidth]{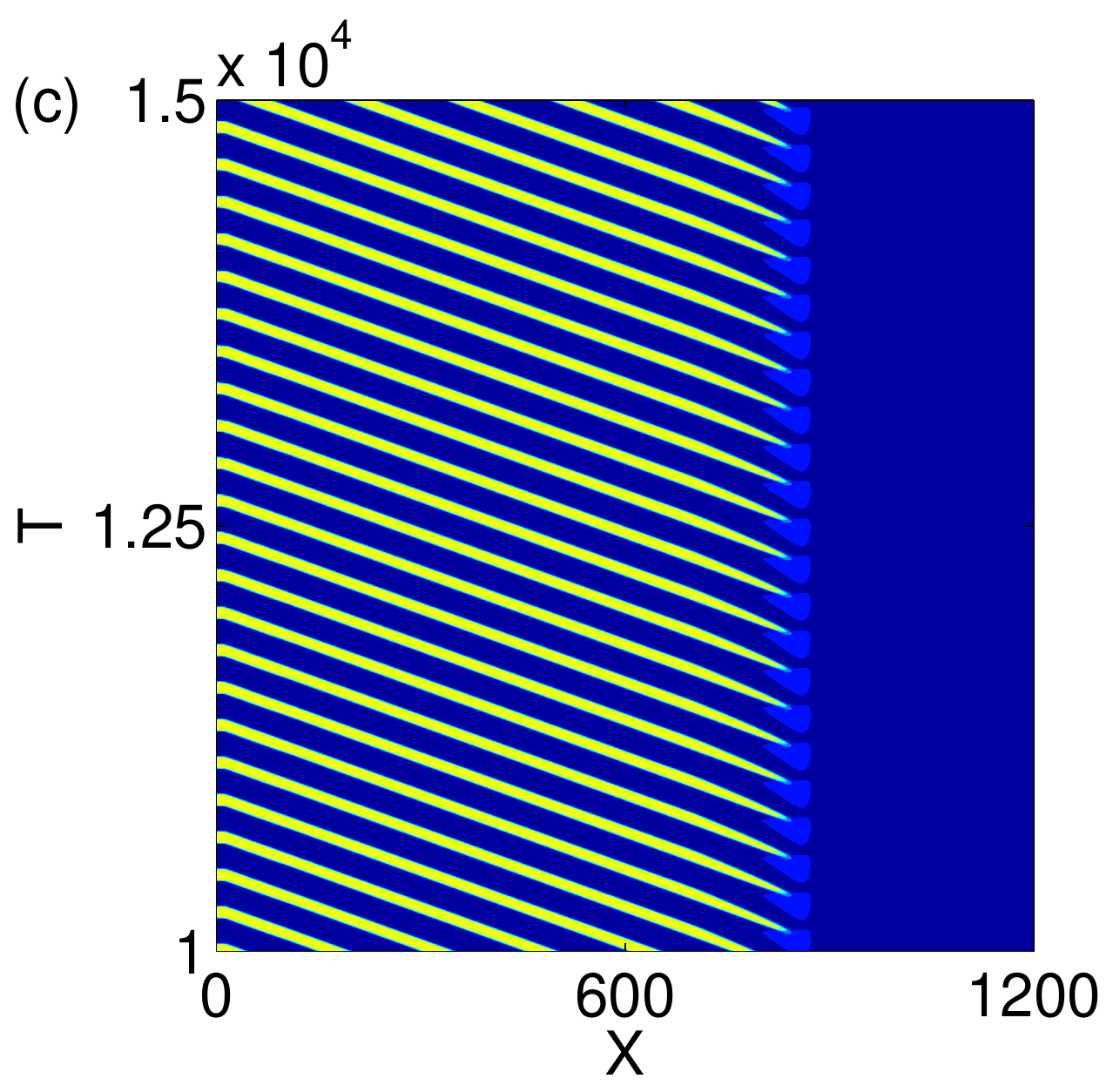}
\caption{\label{fig:spacetimehom}Spacetime plots of solutions on homogeneous substrates ($\rho=0$) at different transfer velocities, obtained from 1D simulations: (a) $V=0.385$ (b) $V=0.500$ (c) $V=0.700$. Only the monolayer density $\Gamma$ is shown. Blue (dark) indicates the liquid expanded phase, while yellow (bright) corresponds to liquid condensed parts of the monolayer.}
\end{figure}
For each value of $\rho$, one can find a range of transfer velocities $V$ for which the surfactant monolayer is deposited in regular patterns of periodically alternating stripes of liquid expanded and liquid condensed material. The wavelength of the patterns, as well as the ratio of the widths of the LE and LC domains depend on $V$.
The patterns consist of travelling domains, which are created at the contact line and then advected by the moving substrate (see figure \ref{fig:spacetimehom} for a spacetime plot of patterns evolving on a homogeneous substrate as obtained from one dimensional simulations). Away from the contact line, the surfactant domains are subject to coarsening. The mobility of the surfactants during this process is dependent on the height of the liquid film underneath (see equation \eqref{evolgamma}). %Therefore, the coarsening, which is certainly taking place in real experiments until the monolayer has completely dried, is slightly exaggerated in the theory due to the existence of the precursor.
\begin{figure}
\includegraphics[width=.5\textwidth]{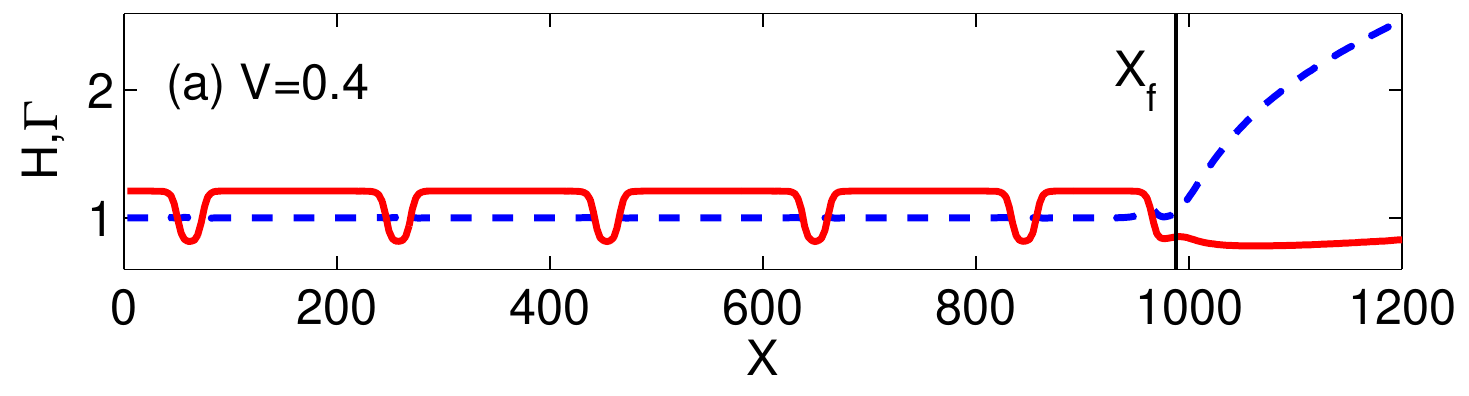}
\includegraphics[width=.5\textwidth]{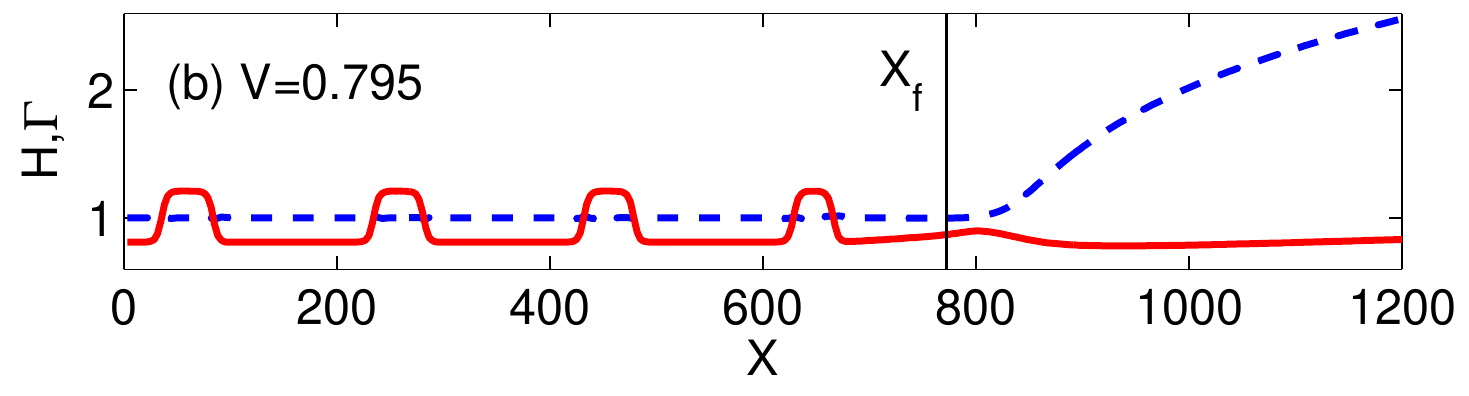}
\caption{\label{fig:snaps_hom}Patterns of equal wavelength, $\lambda_0\approx 200$, obtained from 1D simulations at different velocities $V$ on a homogeneous substrate: (a) $V=0.4$ and (b) $V=0.795$. The red graph (solid line) corresponds to monolayer density $\Gamma$, while the blue graph (dashed line) represent the height profile $H$. The patterns are almost inverse to each other. The solid vertical lines indicate the location $X_\mathrm{f}$ where new domains are perpetually created.}
\end{figure}
We start with a systematic investigation of the synchronization behavior of the one dimensional system. To isolate the effect of the prepatterned substrate on the time-periodic \emph{formation} of surfactant domains from the coarsening, we first determine the location $X_\mathrm{f}$ behind the contact line, where the domains are created (see figure \ref{fig:snaps_hom}). Note, that this location depends on the transfer velocity $V$, since the meniscus is more and more elongated with growing $V$. Then $s(T):=\Gamma(X_\mathrm{f},T)$ is sampled as a time series. By use of a standard procedure from signal analysis, one can unambiguously define an instantaneous phase $\phi_s(T)$ of the ``signal'' $s(T)$. To this end, one constructs the analytic signal $\zeta(T)=s(T)+\mathrm{i}s_H(T)=:C(T)\exp(\mathrm{i}\phi_s(T))$, where $s_H(T)$ is the Hilbert transform of $s(T)$~\cite{PRKsync}. Finally, we obtain the frequency of the domain formation as $\omega=\langle \mathrm{d}\phi_s/\mathrm{d}T\rangle$, where $\langle\dots\rangle$ denotes the time average. This frequency can then be compared to the time periodic forcing frequency $\Omega=2\pi V / L_\mathrm{p}$ due to the prepatterned substrate moving underneath the position $X_\mathrm{f}$.
\subsection{Choice of parameters}
The substrate-liquid interaction is quantified by the Hamaker constant $a_3$. Here, we use a typical value $a_3=0.5/(12\pi)\cdot 10^{-20}\,\mathrm{J}$ and assume a precursor height of $h_p\approx 5.8\,\mathrm{nm}$, yielding the dimensionless Hamaker constant $A=6.5\cdot10^{-2}$.

The critical number density of the first order phase transition between the LE and the LC phase is taken to be $\gamma_\mathrm{cr}=1.539\cdot10^{18}\,m^{-2}$. We assume the surface tension of water covered with a monolayer of density $\gamma_\mathrm{cr}$ to be $10\%$ lower than the surface tension in absence of any surfactant: $\sigma_0=0.9\,\sigma_\mathrm{abs}=65.48\,\mathrm{mN/m}$. Then, the experimental lateral pressure isotherm obtained for DPPC at $20\,^\circ \mathrm{C}$ is reasonably well approximated by using the parameters $M_1=50$ and $M_2=0.2$ in the equations \eqref{fhom} and \eqref{P_hom}. Estimating the surfactant domain line tension $\kappa=5\cdot 10^{-52}\, \mathrm{Jm^4}$, we obtain the time and length scales $l_0=134.49\,\mathrm{nm}$ and $t_0=25.23\,\mathrm{\mu s}$. This yields the smallness parameter $\epsilon=4.31\cdot10^{-2}$.

The evaporation parameters used in our simulation are given by $\mathrm{Ev}=0.45$ and $\mu_\mathrm{v}=5\cdot 10^{-4}$, amounting to a rate constant $Q_\mathrm{e}=4.93\cdot 10^{-9}\, \mathrm{m}^2\mathrm{s} / \mathrm{kg}$.

 The SMC coupling constant is set to $B=1500$, corresponding to a lateral pressure shift $\Delta p_\mathrm{lat}=S(H_\mathrm{p})\,\sigma_0=-3A B \sigma_0/10\approx -3.60\,\mathrm{mN/m}$ caused by SMC. Thus, the coexistence lateral pressure at the substrate is roughly $5\%$ lower than for the floating monolayer.

\subsection{Homogeneous substrates}
\begin{figure}
\includegraphics[width=.5\textwidth]{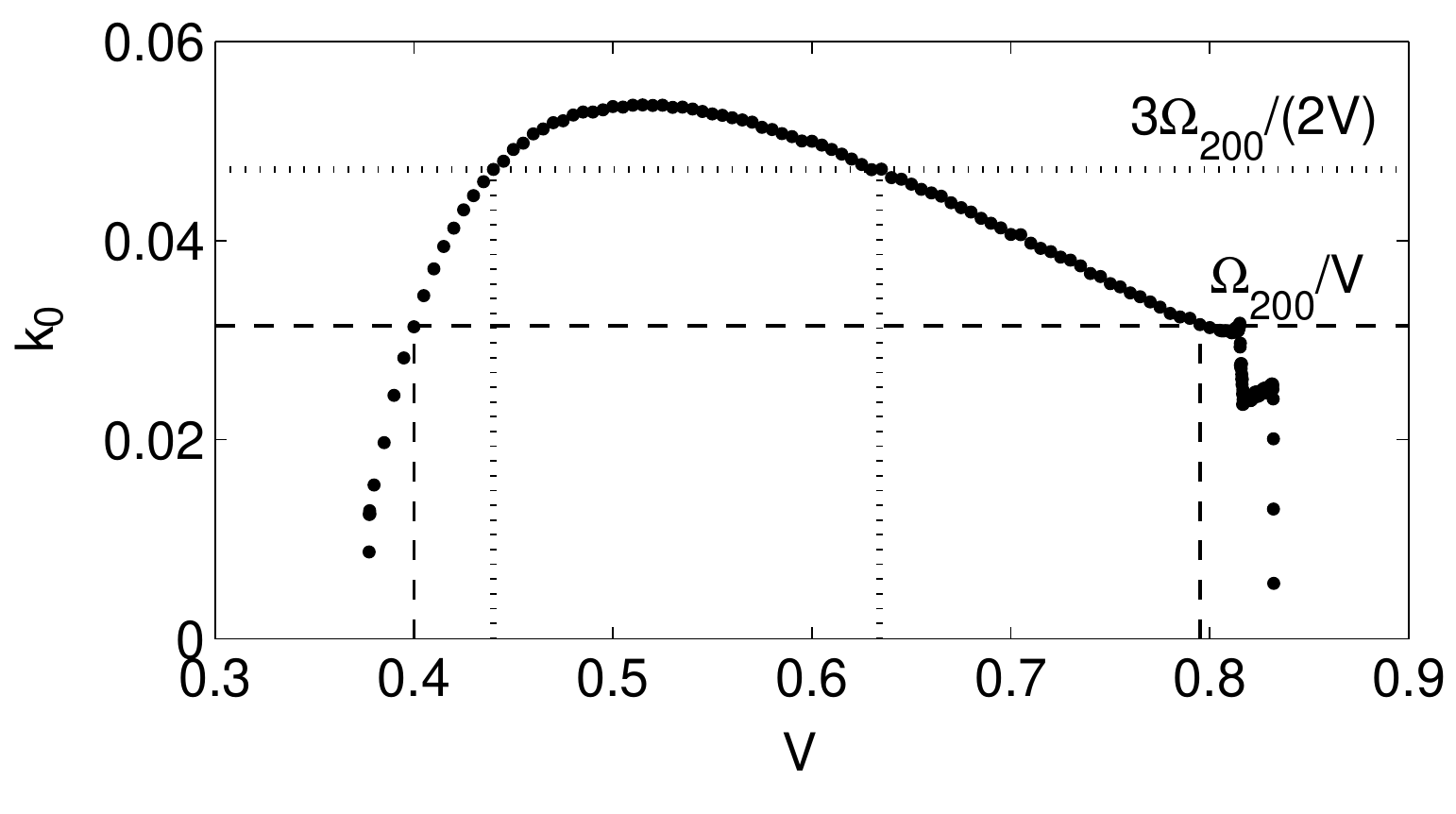}
\caption{\label{fig:freqvel_nopat}Wavenumber $k_0$ of patterns transfered onto homogeneous substrates against transfer velocity $V$, as obtained from 1D computations. The curve is nonmonotonous, so that the same $k_0$ can be obtained for high and for low $V$. The origins of the Arnold tongues in figure \ref{fig:tongues} can be determined from the intersection points of the curve with horizontals at the corresponding $n:m$ ratio. This is shown here for $1:1$ and $3:2$ synchronization with $L_\mathrm{p}=200$.}
\end{figure}

On a homogeneous substrate, $\rho=0$, spatiotemporal oscillations are found in a velocity range $0.38\lesssim V \lesssim 0.83$. Since the corresponding wavenumbers and frequencies are intrinsic quantities of the system, they are denoted by $k_0$ and $\omega_0$, respectively, in order to distinguish them from the wavenumbers $k$ and frequencies $\omega$ obtained on prepatterned substrates. Figure \ref{fig:freqvel_nopat} shows $k_0=\omega_0/V$ as a function of transfer velocity, obtained from one dimensional calculations. Note, that $k_0(V)$ is non-monotonous, yielding pairs of patterns with equal $k$ obtained at different velocities. The two patterns comprising such a pair are roughly inverse to each other, but differ in detail. This is shown in figure \ref{fig:snaps_hom} where two patterns of equal wavelength, $\lambda_0\approx 200$, exhibit different structure: The LE domains in figure \ref{fig:snaps_hom} (a) are narrower than the LC domains in figure \ref{fig:snaps_hom} (b). As will be shown in the next paragraph, these different types of patterns will react slightly different to the periodic forcing caused by prepatterned substrates. 

For most velocities, the two dimensional system behaves analogously, yielding stripes parallel to the contact line, corresponding to the alternating domains of the one-dimensional case. These solutions are symmetric with respect to translations along the $Y$-axis, that is, the system behaves just as expected from the results obtained in the one dimensional case. However, for values of $V$ chosen close to the upper or lower bound of the velocity range where patterns are obtained, the stripe formation is not stable. Instead, after a few stripes are produced, the regularity is destroyed and disordered domains are created. Remarkably, this disordered state is only transient for velocities chosen from the range $V\lesssim0.47$ and marks an orientation transition from parallel to perpendicular stripes~\cite{KGFC_Langmuir_10}. 
\subsection{Prepatterned substrates}
\begin{figure}
\includegraphics[width=.5\textwidth]{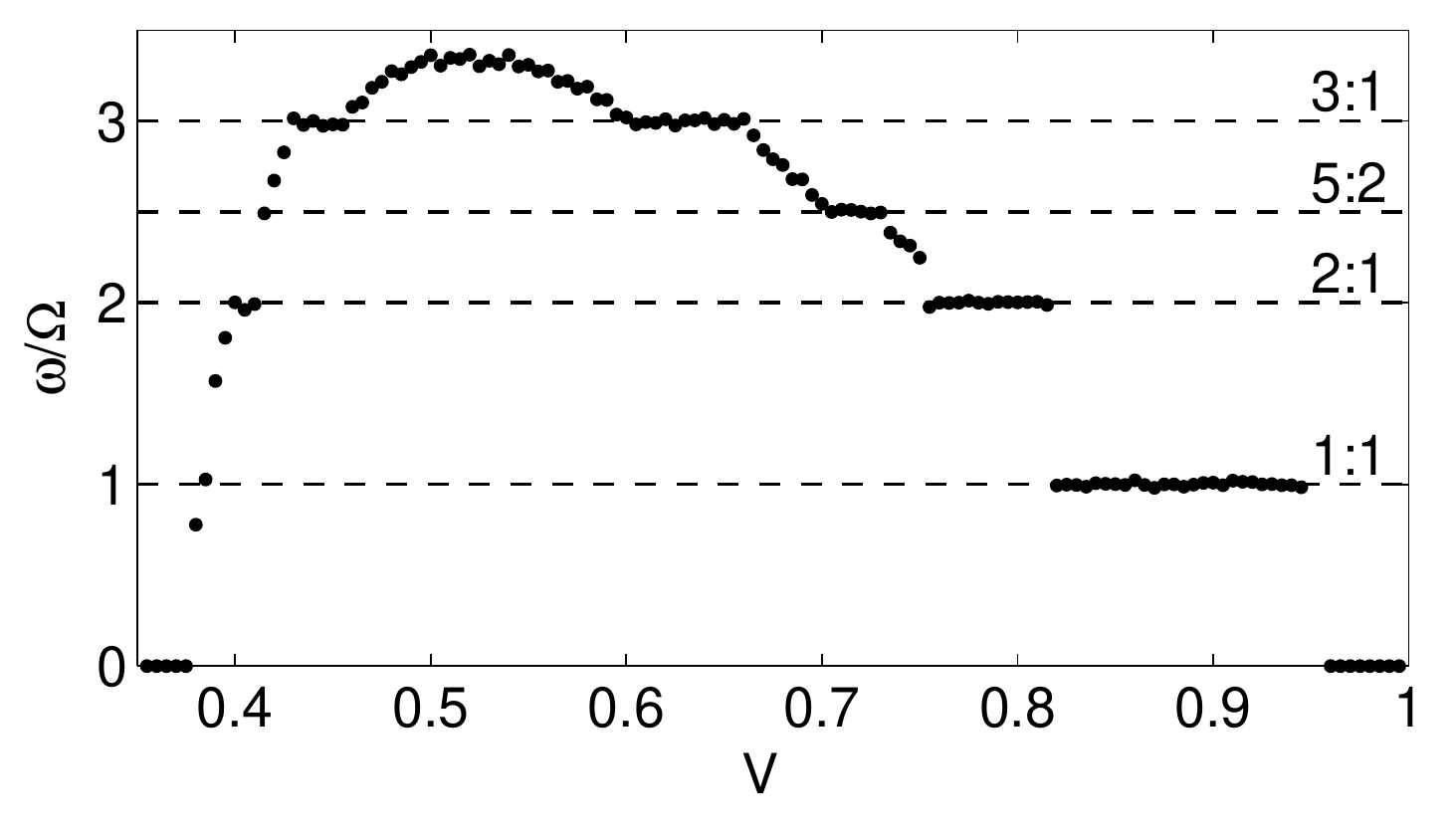}
\caption{\label{fig:freqvel_LA400SA0007}Synchronization plateaus obtained from 1D simulations of transfer onto a prepatterned substrate with $L_\mathrm{p}=400$ and $\rho=0.006$.}
\end{figure}
For $\rho>0$ the wettability contrast of the moving substrate leads to an oscillation of the meniscus, going along with a change of the wavelength of the produced pattern. Within certain velocity ranges, the pattern synchronizes with the substrate, yielding commensurate frequency ratios $\omega/\Omega=n:m$ with integer $n,m$. This can be seen from the pronounced plateaus shown in figure \ref{fig:freqvel_LA400SA0007}. It is also noteworthy, that the overall velocity range where patterns are obtained is extended beyond $V=0.83$, the upper limit on homogeneous substrates.
\begin{figure}
\includegraphics[width=.5\textwidth]{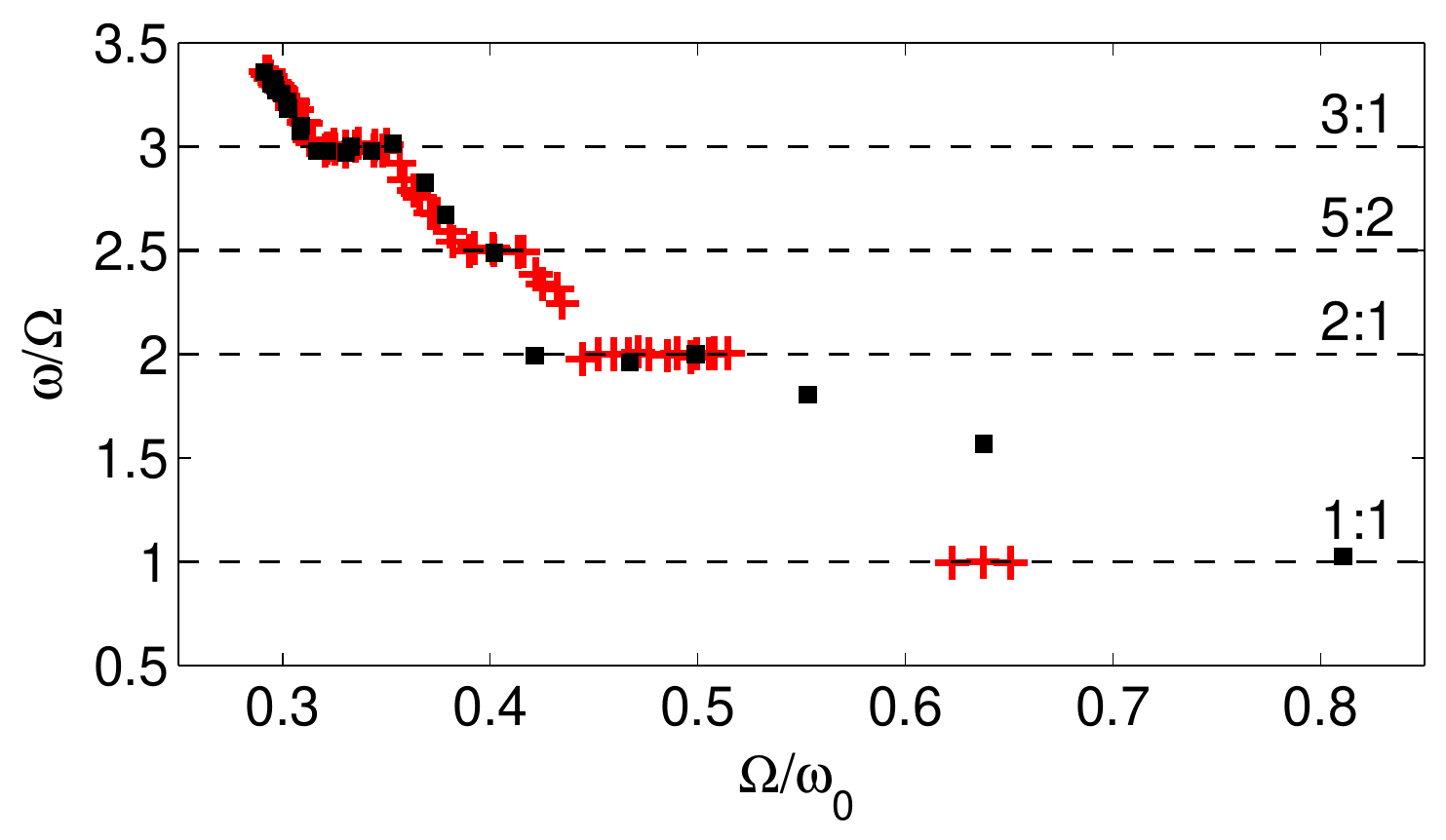}
\caption{\label{fig:freqom_LA400SA0007}Staircase obtained from one dimensional simulations using $L_\mathrm{p}=400$ and $\rho=0.006$, after elimination of the transfer velocity in favor of the natural frequency scale $\Omega/\omega_0$. Red crosses indicate patterns obtained for velocities higher than $0.52$, black squares correspond to lower $V$.}
\end{figure}
As an alternative to the representation used in figure \ref{fig:freqvel_LA400SA0007}, one can also use the ratio $\Omega/\omega_0$ as the abscissa, thereby directly relating the observed frequencies to the natural frequencies of the system without referring to the transfer velocity $V$. In the case considered here, this approach has two disadvantages. Firstly, velocities $V>0.83$, where patterns are only obtained on prepatterned substrates, could not be included in the diagrams, since for them $\omega_0=0$. Secondly, as has been mentioned in explanation of figures \ref{fig:freqvel_nopat} and \ref{fig:snaps_hom}, patterns obtained for higher $V$ differ from those obtained for lower $V$, even if they have exactly the same period. Due to these differences, the synchronization plateaus of the low and the high transfer velocity regimes do not match perfectly. Instead, one obtains two staircases, one for low and one for high $V$, as is shown in figure \ref{fig:freqom_LA400SA0007}. Therefore, we will keep the $V$ representation throughout the remainder of this article.
\begin{figure}
\includegraphics[width=.5\textwidth]{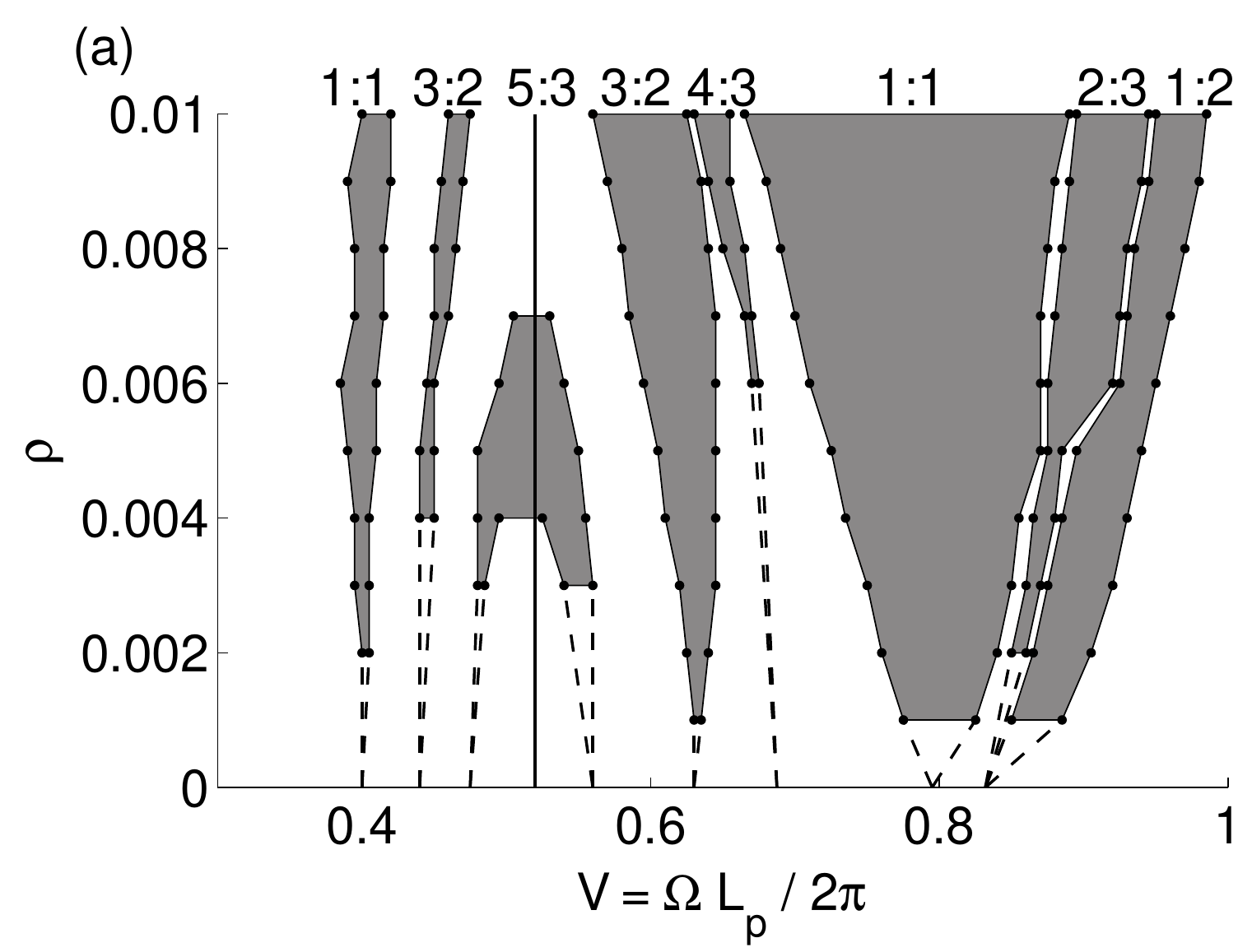}
\includegraphics[width=.5\textwidth]{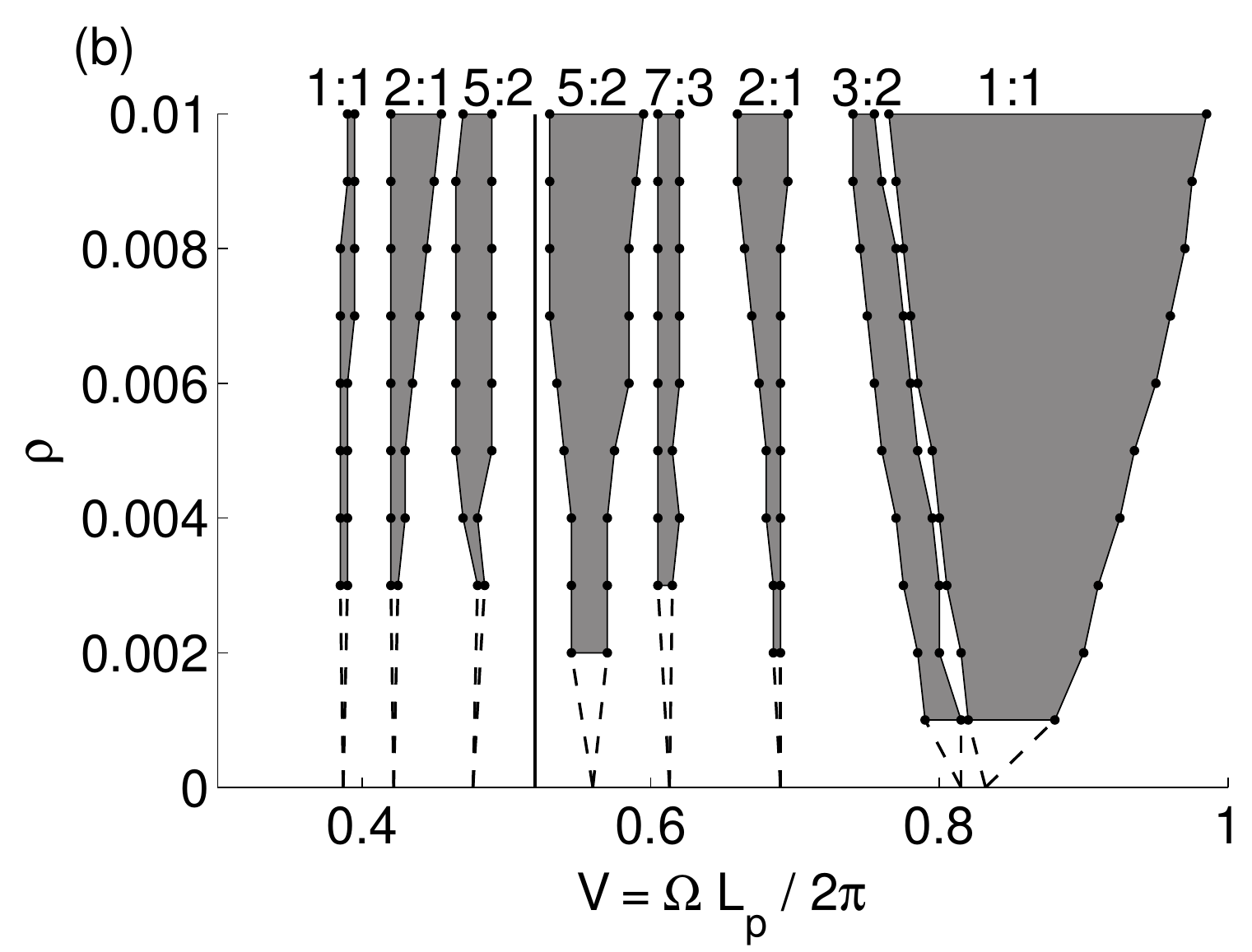}
\includegraphics[width=.5\textwidth]{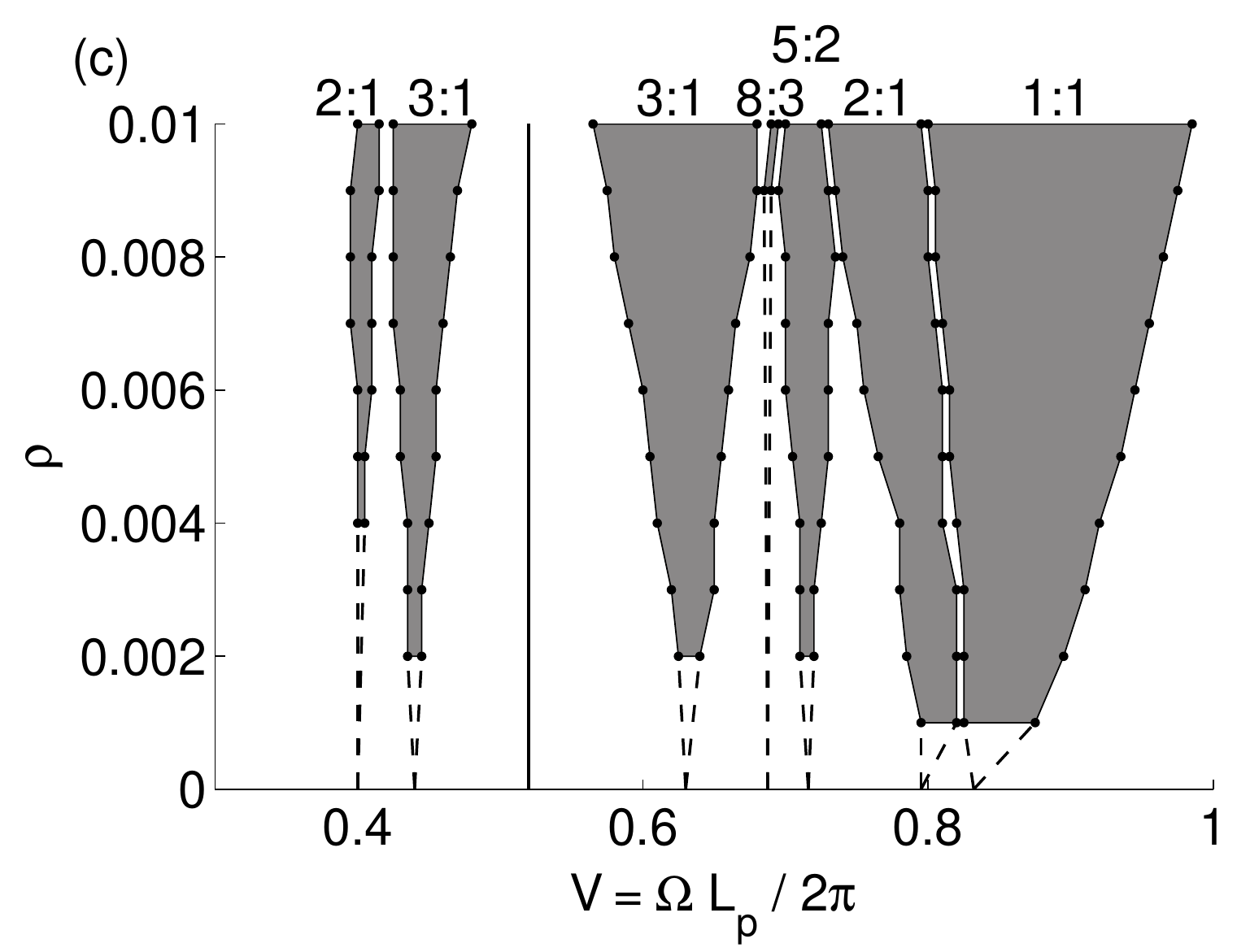}
\caption{\label{fig:tongues}Arnold tongues resulting from one dimensional computations for prepatterned substrate with (a) $L_\mathrm{p}=200$, (b) $L_\mathrm{p}=300$, and (c) $L_\mathrm{p}=400$. The dashed lines show how the Arnold tongues are expected to go to $\rho=0$. The vertical solid line indicates the velocity corresponding to the maximum of the intrinsic frequency $\omega_0$.}
\end{figure}
Naturally, the frequency plateaus broaden with increasing amplitude of the wettability contrast $\rho$. This dependence can be visualized in a $\rho$-$V$ diagram, where the synchronization regions are drawn for different values of $\rho$, yielding the famous Arnold tongues~\cite{PRKsync}. Figure \ref{fig:tongues} shows these diagrams for the three different considered wavelengths $L_\mathrm{p}$.
Generally, an $n:m$ Arnold tongue meets the $V$-axis at the velocity corresponding to the $n:m$ ratio of the intrinsic wavenumber $k_0$ and the forcing wavenumber $k_\Omega$. This velocity can be easily found by looking at the intersections of a horizontal line at $k_0=k_\Omega n/m$ with the curve $k_0(V)$. This is shown in figure \ref{fig:freqvel_nopat} for the examples of $1:1$ and $3:2$ synchronization with $L_\mathrm{p}=200$.
\begin{figure}
\includegraphics[width=.5\textwidth]{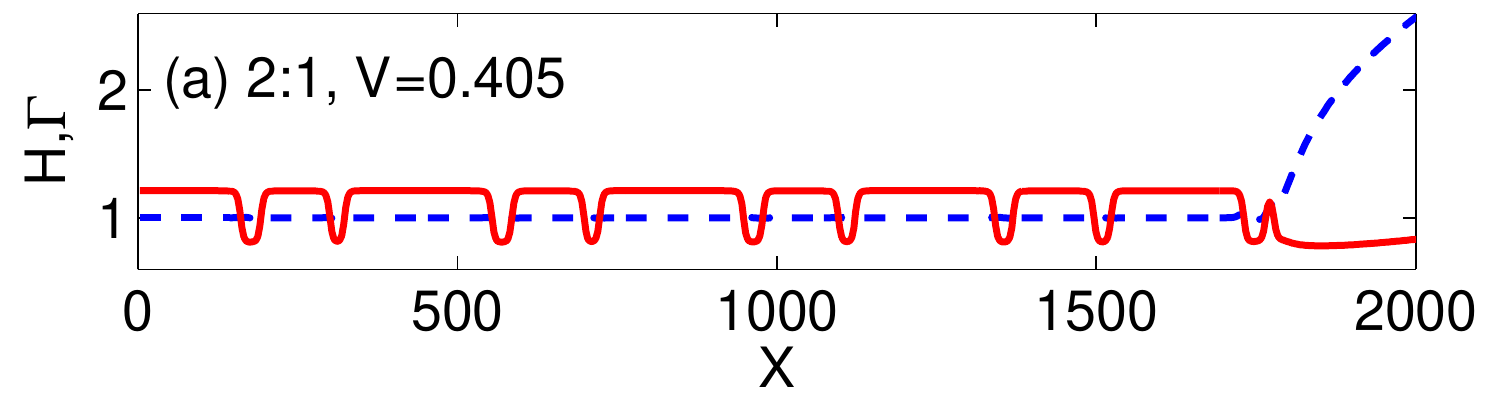}
\includegraphics[width=.5\textwidth]{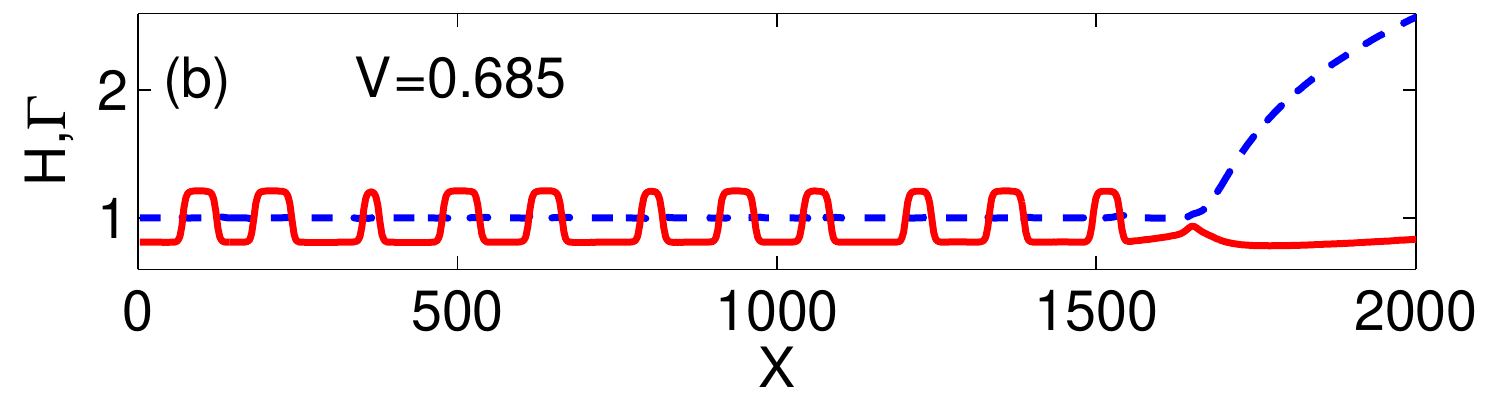}
\includegraphics[width=.5\textwidth]{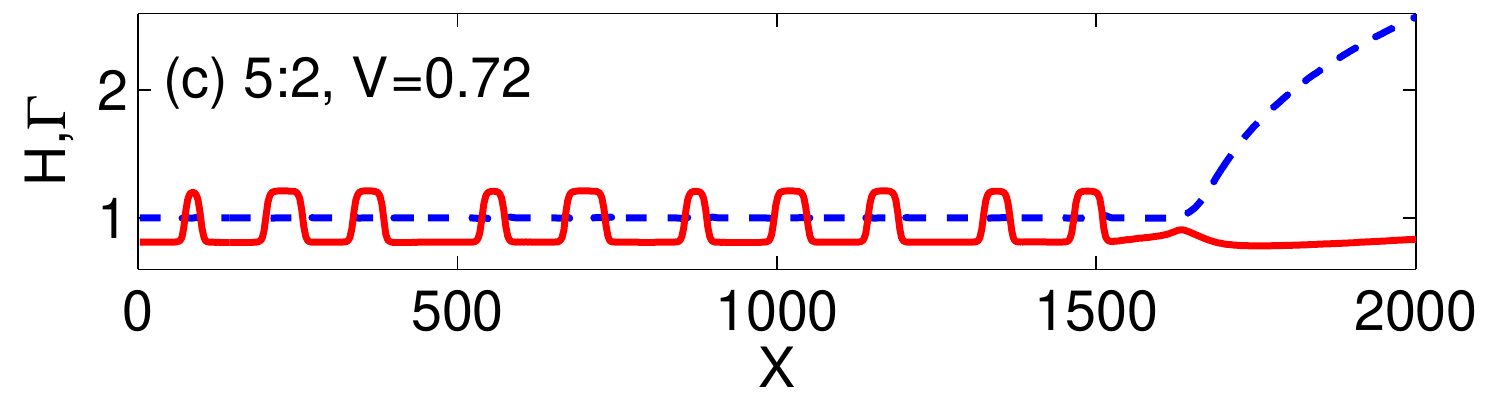}
\includegraphics[width=.5\textwidth]{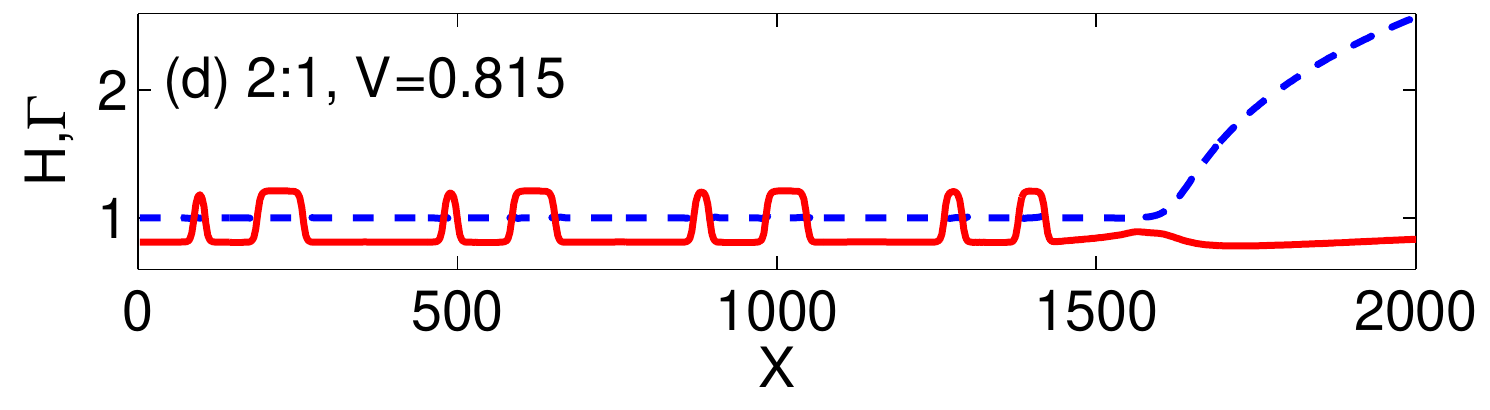}
\includegraphics[width=.5\textwidth]{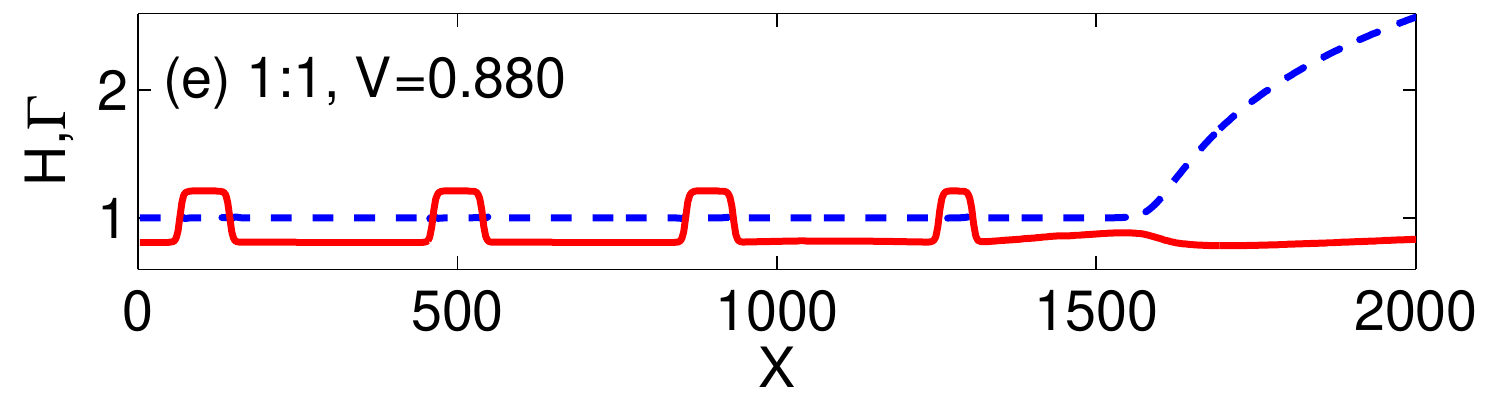}
\caption{\label{fig:snap_LA400SA0005}Patterns in the 1D system, obtained during transfer with five different velocities on a prepatterned substrate with $L_\mathrm{p}=400$, $\rho=0.005$. The velocities in (a) and (c-f) are chosen from different $n:m$ synchronization regions, while (b) shows a pattern obtained at a velocity from between the right $3:1$ and $2:1$ regions (compare \ref{fig:tongues} (a)).}
\end{figure}
In comparison to the case of homogeneous substrates (see figures \ref{fig:spacetimehom} and \ref{fig:snaps_hom}), transfer onto prepatterned substrates yields patterns of higher complexity. Figure \ref{fig:snap_LA400SA0005} shows snapshots of transfer using a periodic wettability contrast of wavelength $L_\mathrm{p}=400$ and amplitude $\rho=0.005$. One has to distinguish two cases. For velocities which belong to one of the synchronization plateaus, the obtained pattern is periodic. Figure \ref{fig:snap_LA400SA0005} a) for example shows a periodic train of domain pairs produced by $2:1$ synchronization. If the chosen velocity instead lies outside the plateaus, a nonperiodic train of domains is obtained (see figure \ref{fig:snap_LA400SA0005} b)). By eye, it is generally difficult to distinguish a non-synchronous pattern from a synchronous pattern with higher values of $n$ and $m$ as can be seen by comparison of figures \ref{fig:snap_LA400SA0005} b) and \ref{fig:snap_LA400SA0005} c). For the system under consideration, this difficulty is even enhanced since, as has been mentioned above, the domains are subject to coarsening from the moment they detach from the contact line. The coarsening is obvious in figure \ref{fig:snap_LA400SA0005} d), where the leftmost, that is oldest, pair of domains already differs visibly from the rightmost pair which has just been formed at the meniscus.
\begin{figure}
\includegraphics[width=.23\textwidth]{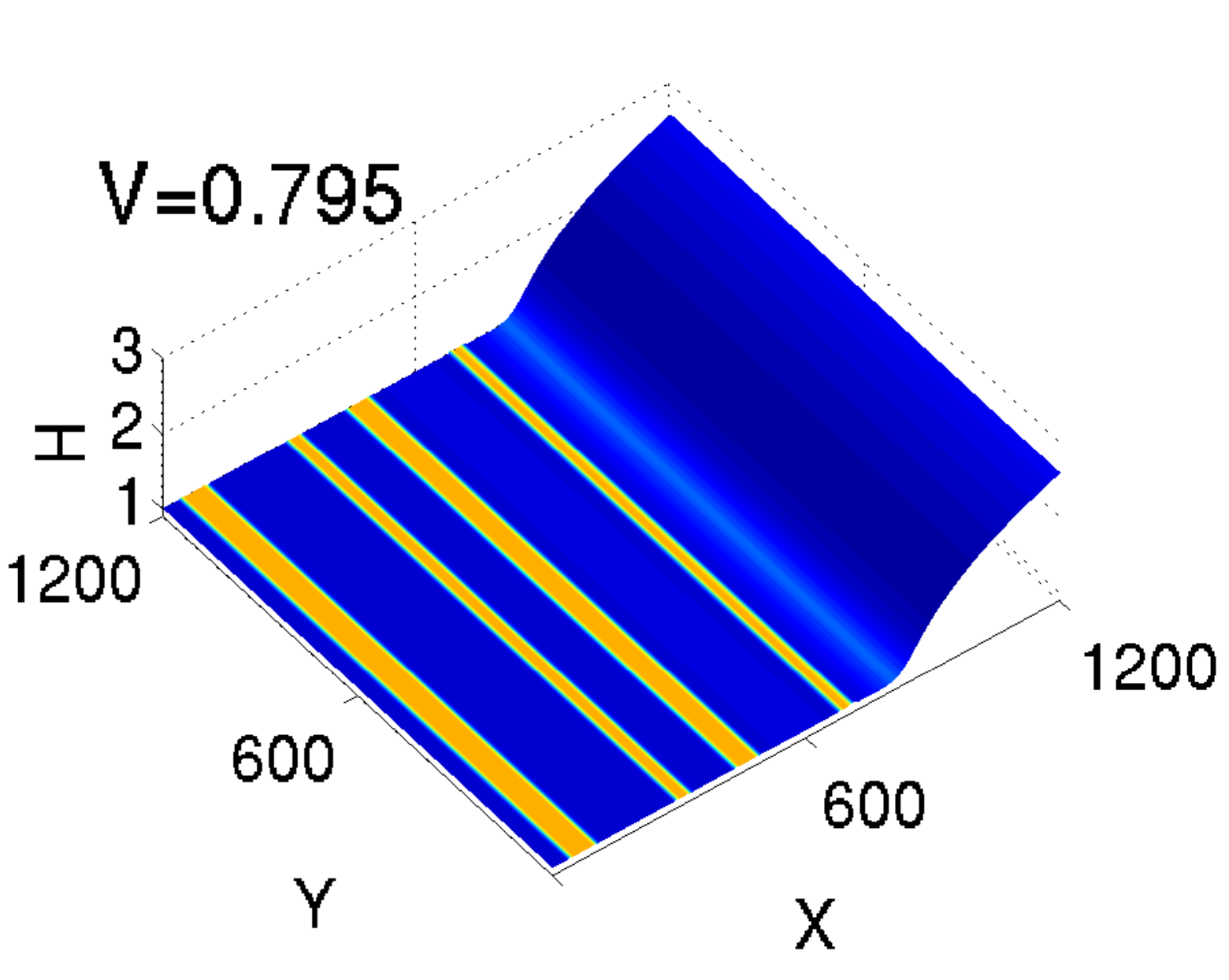}
\includegraphics[width=.23\textwidth]{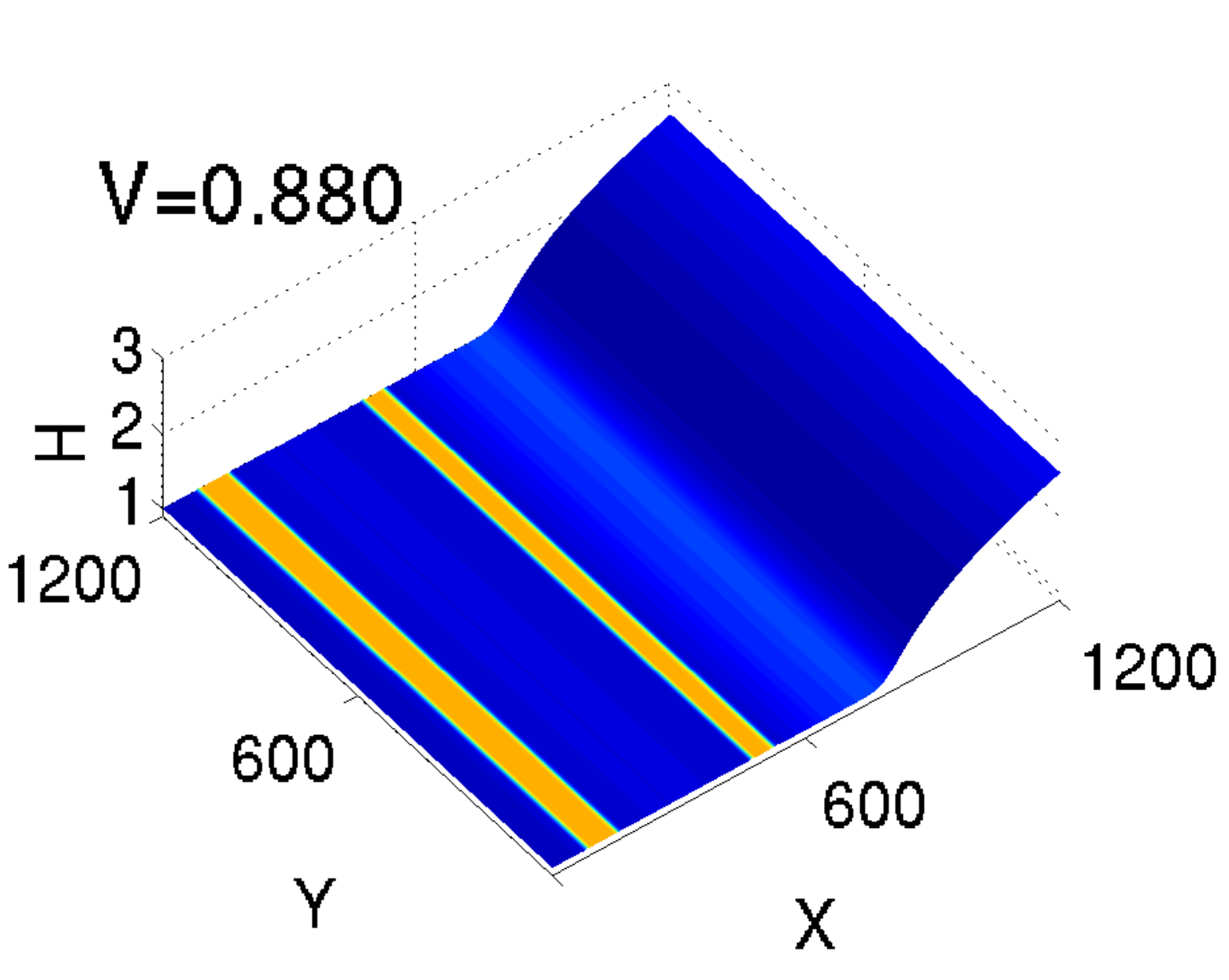}
\includegraphics[width=.34\textwidth]{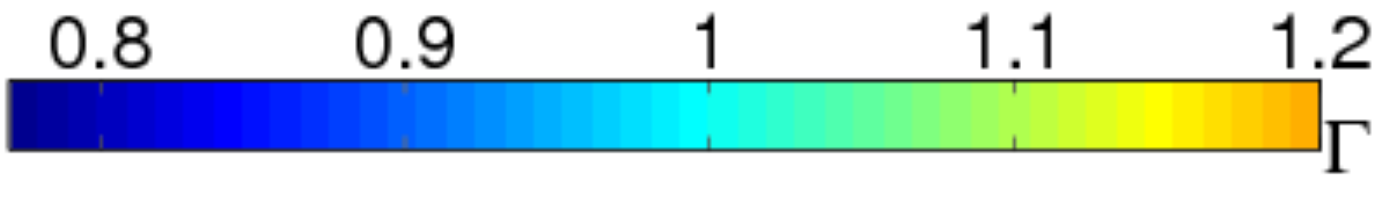}
\caption{\label{fig:prep2D}Two-dimensional patterns obtained on a prepatterned substrate with $L_\mathrm{p}=400$ for two different transfer velocities $V$. The height information corresponds to the film profile $H$, while the monolayer density $\Gamma$ is color-coded.} 
\end{figure}

Like on homogeneous substrates, in the case of two spatial dimensions, stripe patterns arranged parallel to the contact line are obtained analogously to the domains found in the one dimensional case. The periodicity of these patterns matches the one dimensional solutions. Figure \ref{fig:prep2D} shows two snapshots of numerical simulations for a prepattern of period $L_\mathrm{p}$, 
However, as has been mentioned above in the discussion of transfer onto homogeneous substrates, there exists a velocity range, where the formation of stripes arranged parallel to the contact line competes with a stationary pattern of perpendicular stripes. Due to the presence of the prepattern, this competition gives rise to the formation of new intermediate patterns, which are neither stripes parallel nor stripes perpendicular to the contact line and which are not observed on homogeneous substrates. This is explained in detail in the next subsection.
\subsection{The orientation transition}
\begin{figure}
\begin{tabular}{cc}
\includegraphics[width=.23\textwidth]{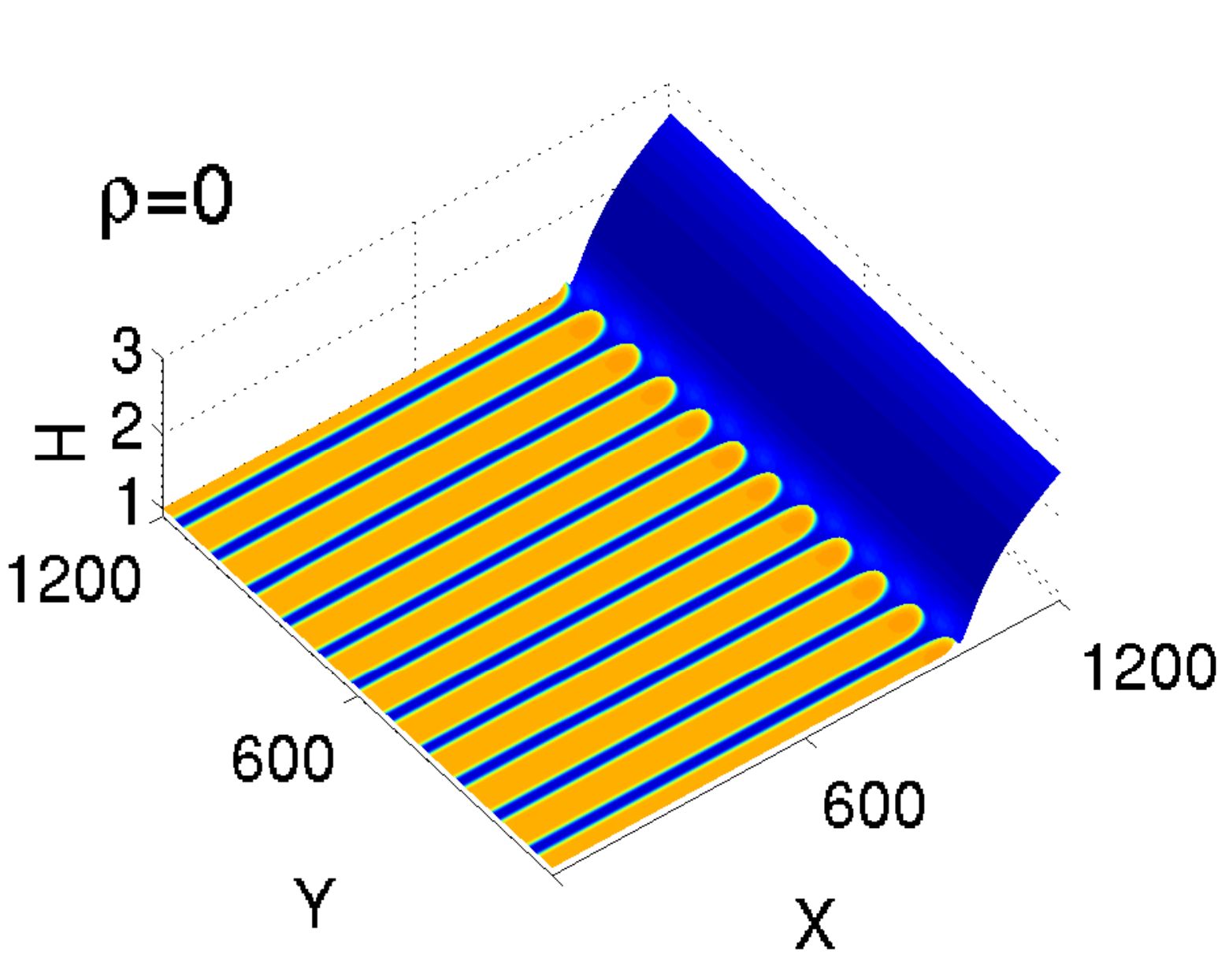} & \includegraphics[width=.23\textwidth]{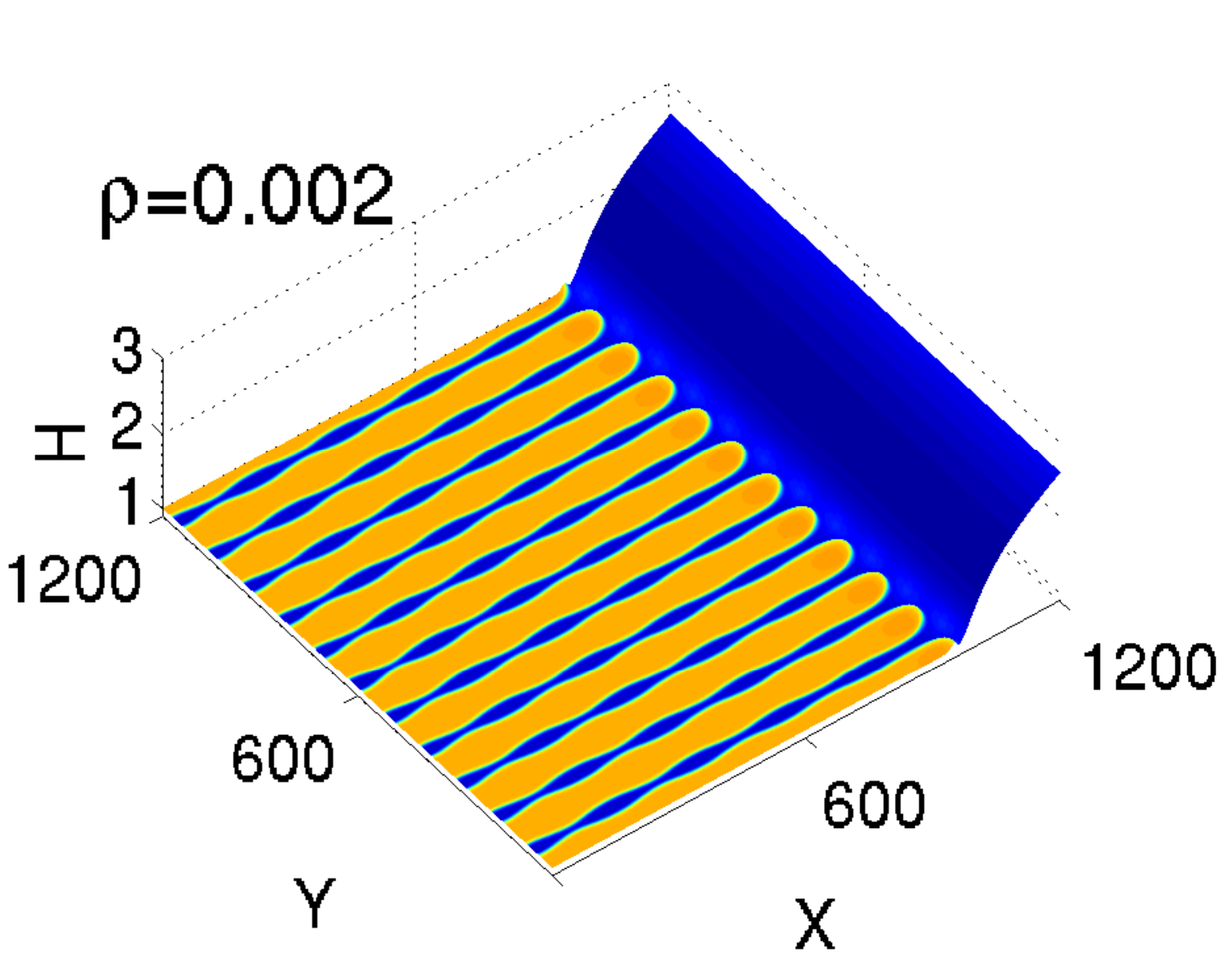} \\
\includegraphics[width=.23\textwidth]{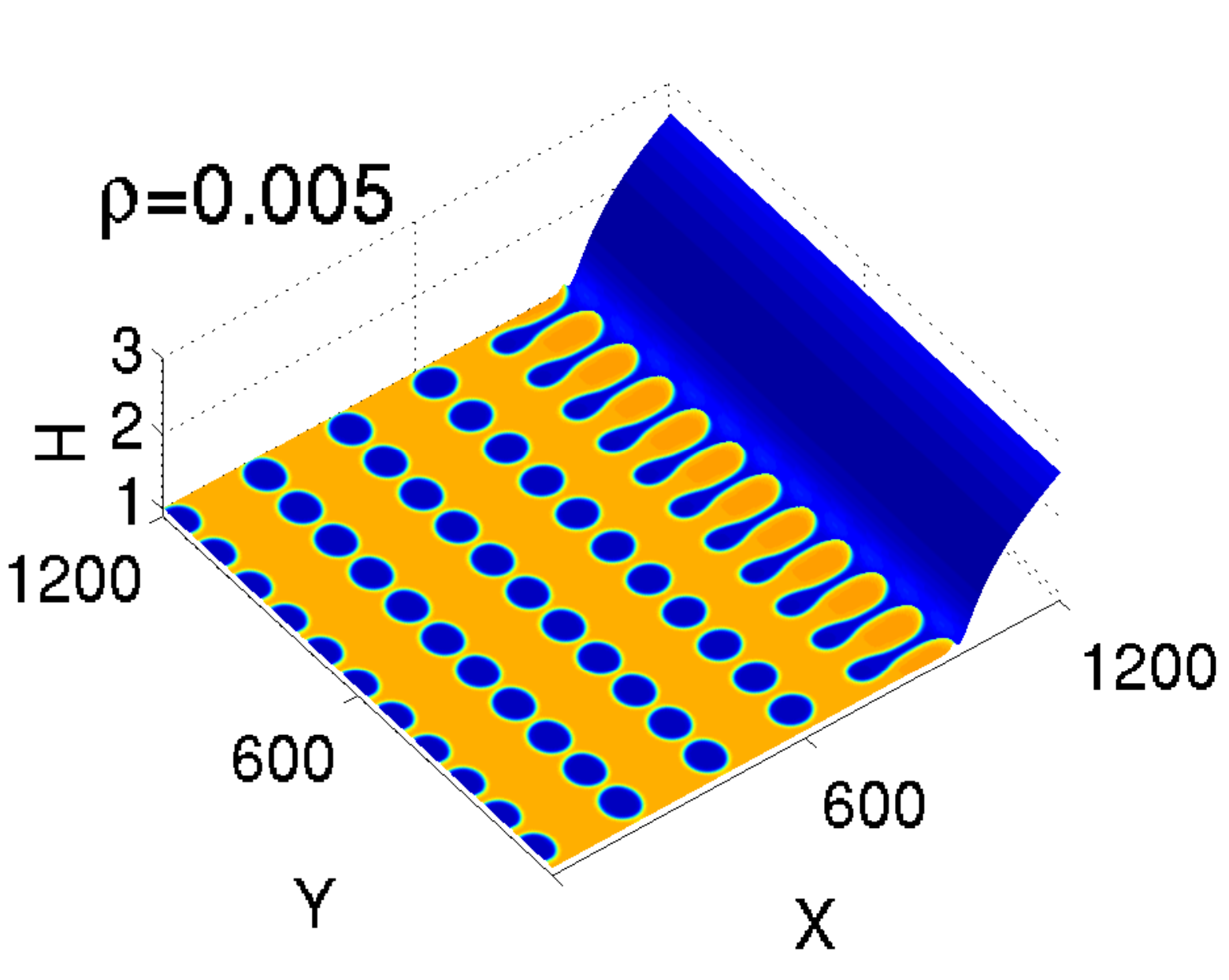} &
\includegraphics[width=.23\textwidth]{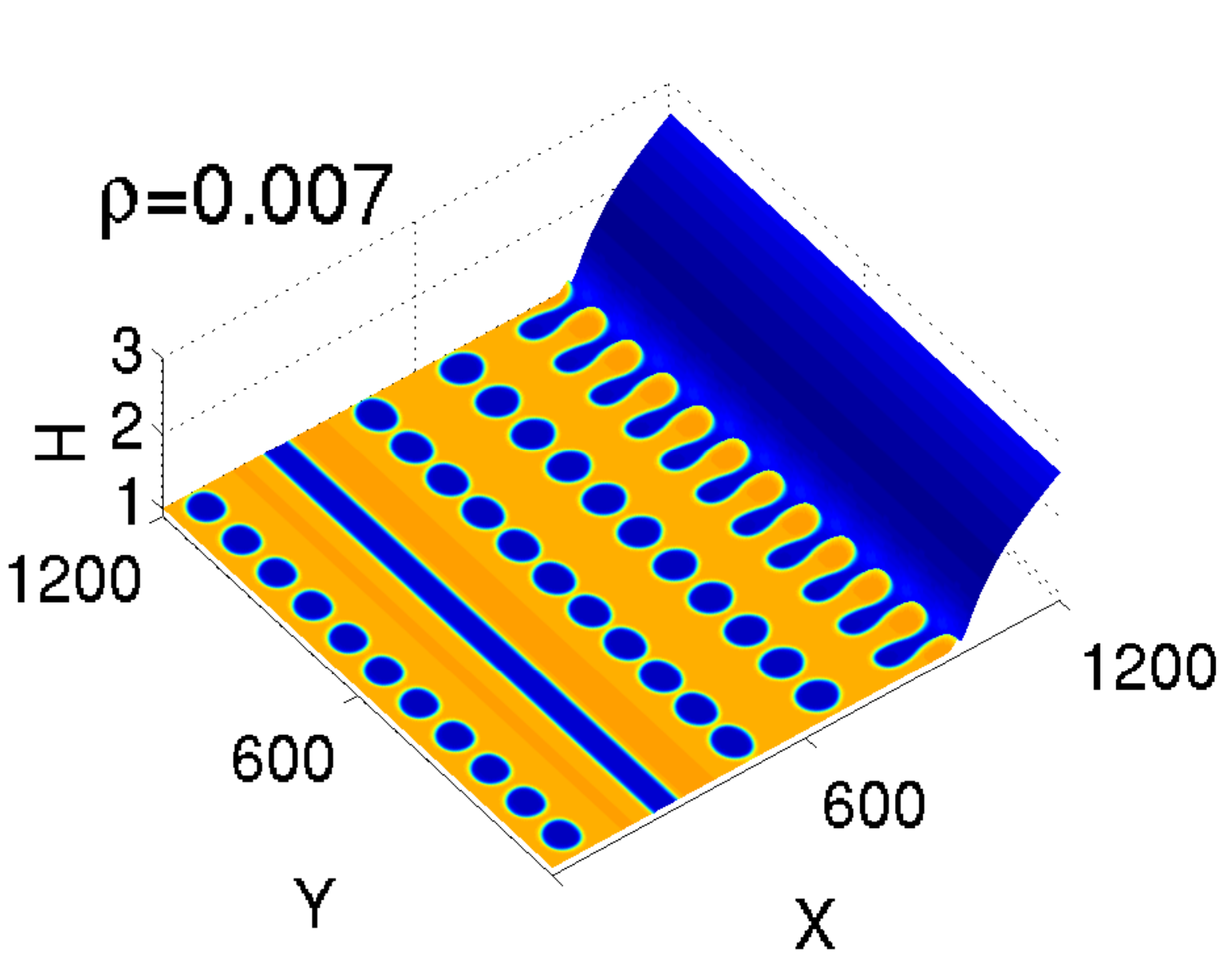} \\
\includegraphics[width=.23\textwidth]{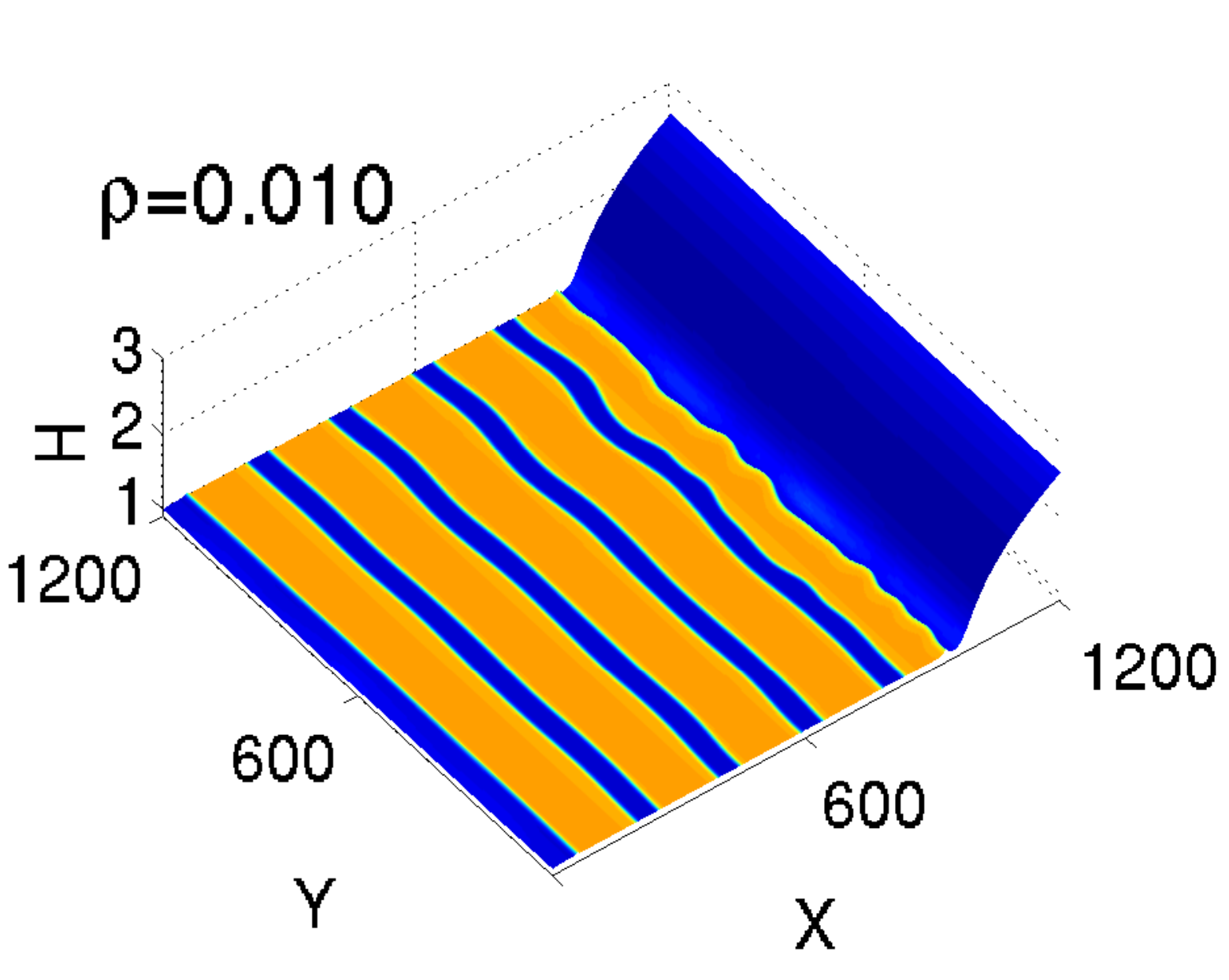} & \includegraphics[width=.060\textwidth]{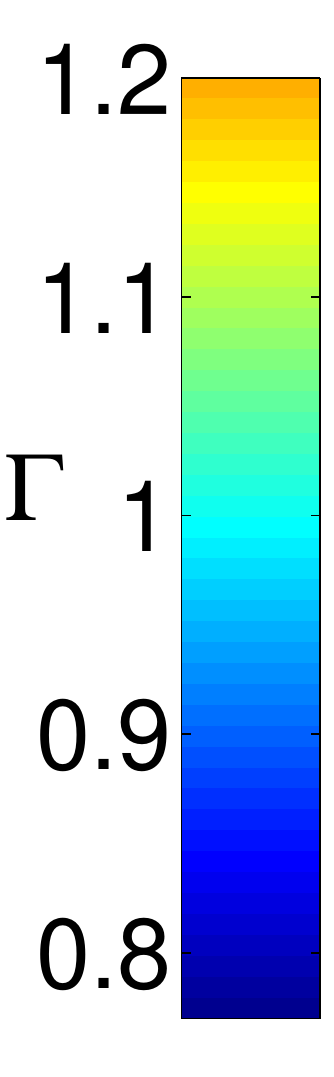}
\end{tabular}
\caption{\label{fig:rhoinc}Two dimensional solutions in the velocity range of the parallel-perpendicular orientation transition for various wetting contrasts $\rho$.}
\end{figure}
During transfer onto homogeneous substrates, a transition from stripes parallel to the contact line to perpendicular stripes was observed for velocities $0.38\lesssim V\lesssim0.47$. On prepatterned substrates, these perpendicular stripes are stable only for small wetting contrast $\rho$. To illustrate this, we consider the solution obtained for $\rho=0,V=0.44$, that is a stable, stationary pattern of stripes perpendicular to the contact line obtained on a homogeneous substrate. This solution is then used as the initial condition for computations with non-zero $\rho$. In these simulations, the wetting contrast is not activated instantaneously across the whole domain. Instead, the region with $\rho>0$ is moving in from the right boundary with the transfer velocity $V$, thereby mimicking transfer onto a substrate, which is divided into an homogeneous and a prepatterned part. This means, that the monolayer is first transfered onto a homogeneous substrate, where it forms stripes perpendicular to the contact line. Then, at $T=0$, the contact line reaches the prepatterned part of the substrate. Figure \ref{fig:rhoinc} shows, how the wetting contrast impacts the initially regular stripe pattern. For small amplitudes such as $\rho=0.002$, the stripes persist but are periodically undulated across the $X$-direction. These undulations travel along with the moving prepattern underneath. For higher values of $\rho$ the stripes break up into well-arranged arrays of  circular LE domains, aligned like beads on a string parallel to the contact line, and advected with the substrate. For even higher amplitudes $\rho\approx0.007$, these linearly arranged LE domains alternate with solid LE stripes. Finally, for very strong wetting contrasts $\rho\gtrsim0.01$, the extrinsic structure of the prepattern is completely enforced upon the system, resulting in a regular $L_\mathrm{p}$-periodic stripe pattern .
\section{Conclusions and outlook}
A theoretical investigation of monolayer transfer onto chemically prepatterned substrates has been presented. In our model, a spatially varying disjoining pressure accounts for the wetting heterogeneity due to the prepattern. Focussing on the case, where the wetting properties change periodically in the direction perpendicular to the contact line, complex periodic patterns were obtained as a result of synchronization effects of various orders. In addition, it has been shown, that a sufficiently strong wetting contrast can be used to inhibit the transition from stripes parallel to the contact line to perpendicular stripes. Also, for moderate wetting contrasts, well-ordered arrays of circular LE domains are obtained, as well as mixed patterns consisting of alternating stripes and circular domains.
These results demonstrate, that even simple prepatterns provide means to create various complex, well-ordered patterns in monolayer transfer systems and therefore open new perspectives for the controlled nanopatterning of surfaces. 

It can be expected, from our findings, that patterns of higher complexity might be obtained by use of more complicated prepatterns which are not necessarily structured periodically. But even using a periodic wetting heterogeneity as is considered in this article, interesting new patterns might emerge, if the substrate is withdrawn under another angle, so that the stripes of the prepattern are no longer parallel to the contact line. These possibilities remain to be elaborated in a joint effort of theoretical and experimental research.

\begin{acknowledgments}
This work was supported by the Deutsche Forschungsgemeinschaft within SRF TRR 61. We thank L.~F.~Chi for interesting and helpful discussions.
\end{acknowledgments}

\begin{thebibliography}{10}%
\makeatletter
\providecommand \@ifxundefined [1]{%
 \ifx #1\undefined \expandafter \@firstoftwo
 \else \expandafter \@secondoftwo
\fi
}%
\providecommand \@ifnum [1]{%
 \ifnum #1\expandafter \@firstoftwo
 \else \expandafter \@secondoftwo
\fi
}%
\providecommand \enquote [1]{``#1''}%
\providecommand \bibnamefont  [1]{#1}%
\providecommand \bibfnamefont [1]{#1}%
\providecommand \citenamefont [1]{#1}%
\providecommand\href[0]{\@sanitize\@href}%
\providecommand\@href[1]{\endgroup\@@startlink{#1}\endgroup\@@href}%
\providecommand\@@href[1]{#1\@@endlink}%
\providecommand \@sanitize [0]{\begingroup\catcode`\&12\catcode`\#12\relax}%
\@ifxundefined \pdfoutput {\@firstoftwo}{%
 \@ifnum{\z@=\pdfoutput}{\@firstoftwo}{\@secondoftwo}%
}{%
 \providecommand\@@startlink[1]{\leavevmode}%
 \providecommand\@@endlink[0]{}%
}{%
 \providecommand\@@startlink[1]{%
  \leavevmode
  \pdfstartlink
   attr{/Border[0 0 1 ]/H/I/C[0 1 1]}%
   user{/Subtype/Link/A<</Type/Action/S/URI/URI(#1)>>}%
  \relax
 }%
 \providecommand\@@endlink[0]{\pdfendlink}%
}%
\providecommand \url  [0]{\begingroup\@sanitize \@url }%
\providecommand \@url [1]{\endgroup\@href {#1}{\urlprefix}}%
\providecommand \urlprefix [0]{URL }%
\providecommand \Eprint[0]{\href }%
\@ifxundefined \urlstyle {%
  \providecommand \doi [1]{doi:\discretionary{}{}{}#1}%
}{%
  \providecommand \doi [0]{doi:\discretionary{}{}{}\begingroup
  \urlstyle{rm}\Url }%
}%
\providecommand \doibase [0]{http://dx.doi.org/}%
\providecommand \Doi[1]{\href{\doibase#1}}%
\providecommand \bibAnnote [3]{%
  \BibitemShut{#1}%
  \begin{quotation}\noindent
    \textsc{Key:}\ #2\\\textsc{Annotation:}\ #3%
  \end{quotation}%
}%
\providecommand \bibAnnoteFile [2]{%
  \IfFileExists{#2}{\bibAnnote {#1} {#2} {\input{#2}}}{}%
}%
\providecommand \typeout [0]{\immediate \write \m@ne }%
\providecommand \selectlanguage [0]{\@gobble}%
\providecommand \bibinfo [0]{\@secondoftwo}%
\providecommand \bibfield [0]{\@secondoftwo}%
\providecommand \translation [1]{[#1]}%
\providecommand \BibitemOpen[0]{}%
\providecommand \bibitemStop [0]{}%
\providecommand \bibitemNoStop [0]{.\EOS\space}%
\providecommand \EOS [0]{\spacefactor3000\relax}%
\providecommand \BibitemShut [1]{\csname bibitem#1\endcsname}%
%</preamble>
\bibitem{GHLL_Science_99}%
  \BibitemOpen
  \bibfield{author}{%
  \bibinfo {author} {\bibfnamefont{H.}~\bibnamefont{Gau}}, \bibinfo {author}
  {\bibfnamefont{S.}~\bibnamefont{Herminghaus}}, \bibinfo {author}
  {\bibfnamefont{P.}~\bibnamefont{Lenz}},\ and\ \bibinfo {author}
  {\bibfnamefont{R.}~\bibnamefont{Lipowsky}},\ }%
  \bibfield{journal}{%
  \bibinfo {journal} {Science}\ }%
  \textbf{\bibinfo {volume} {283}},\ \bibinfo {pages} {46} (\bibinfo {year}
  {1999})
  \bibAnnoteFile{NoStop}{GHLL_Science_99}%
\bibitem{QXY_AdvMat_99}%
  \BibitemOpen
  \bibfield{author}{%
  \bibinfo {author} {\bibfnamefont{D.}~\bibnamefont{Qin}}, \bibinfo {author}
  {\bibfnamefont{Y.}~\bibnamefont{Xia}}, \bibinfo {author}
  {\bibfnamefont{B.}~\bibnamefont{Xu}}, \bibinfo {author}
  {\bibfnamefont{H.}~\bibnamefont{Yang}}, \bibinfo {author}
  {\bibfnamefont{C.}~\bibnamefont{Zhu}},\ and\ \bibinfo {author}
  {\bibfnamefont{G.~M.}\ \bibnamefont{Whitesides}},\ }%
  \bibfield{journal}{%
  \bibinfo {journal} {Adv. Mater.}\ }%
  \textbf{\bibinfo {volume} {11}},\ \bibinfo {pages} {1433} (\bibinfo {year}
  {1999})
  \bibAnnoteFile{NoStop}{QXY_AdvMat_99}%
\bibitem{SOG_AdvMat_02}%
  \BibitemOpen
  \bibfield{author}{%
  \bibinfo {author} {\bibfnamefont{H.}~\bibnamefont{Shimoda}}, \bibinfo
  {author} {\bibfnamefont{S.}~\bibnamefont{Oh}}, \bibinfo {author}
  {\bibfnamefont{H.}~\bibnamefont{Geng}}, \bibinfo {author}
  {\bibfnamefont{R.}~\bibnamefont{Walker}}, \bibinfo {author}
  {\bibfnamefont{X.}~\bibnamefont{Zhang}}, \bibinfo {author}
  {\bibfnamefont{L.}~\bibnamefont{McNeil}},\ and\ \bibinfo {author}
  {\bibfnamefont{O.}~\bibnamefont{Zhou}},\ }%
  \bibfield{journal}{%
  \bibinfo {journal} {Adv. Mater.}\ }%
  \textbf{\bibinfo {volume} {14}},\ \bibinfo {pages} {899} (\bibinfo {year}
  {2002}
  \bibAnnoteFile{NoStop}{SOG_AdvMat_02}%
\bibitem{CGO_PRL_06}%
  \BibitemOpen
  \bibfield{author}{%
  \bibinfo {author} {\bibfnamefont{A.}~\bibnamefont{Checco}}, \bibinfo {author}
  {\bibfnamefont{O.}~\bibnamefont{Gang}},\ and\ \bibinfo {author}
  {\bibfnamefont{B.~M.}\ \bibnamefont{Ocko}},\ }%
  \bibfield{journal}{%
  \bibinfo {journal} {Phys. Rev. Lett.}\ }%
  \textbf{\bibinfo {volume} {96}},\ \bibinfo {pages} {056104} 
  \bibAnnoteFile{NoStop}{CGO_PRL_06}%
\bibitem{GCF_Nature_00}%
  \BibitemOpen
  \bibfield{author}{%
  \bibinfo {author} {\bibfnamefont{M.}~\bibnamefont{{Gleiche}}}, \bibinfo
  {author} {\bibfnamefont{L.~F.}\ \bibnamefont{{Chi}}},\ and\ \bibinfo {author}
  {\bibfnamefont{H.}~\bibnamefont{{Fuchs}}},\ }%
  \bibfield{journal}{%
  \Doi{10.1038/35003149}{\bibinfo {journal} {Nature}}\ }%
  \textbf{\bibinfo {volume} {403}},\ \bibinfo {pages} {173} (\bibinfo {year}
  {2000})%
  \bibAnnoteFile{NoStop}{GCF_Nature_00}%
\bibitem{HJ_NatMater_05}%
  \BibitemOpen
  \bibfield{author}{%
  \bibinfo {author} {\bibfnamefont{J.}~\bibnamefont{Huang}}, \bibinfo {author}
  {\bibfnamefont{F.}~\bibnamefont{Kim}}, \bibinfo {author}
  {\bibfnamefont{A.~R.}\ \bibnamefont{Tao}}, \bibinfo {author}
  {\bibfnamefont{S.}~\bibnamefont{Connor}},\ and\ \bibinfo {author}
  {\bibfnamefont{P.}~\bibnamefont{Yang}},\ }%
  \bibfield{journal}{%
  \bibinfo {journal} {Nature Materials}\ }%
  \textbf{\bibinfo {volume} {4}},\ \bibinfo {pages} {896} (\bibinfo {year}
  {2005})%
  \bibAnnoteFile{NoStop}{HJ_NatMater_05}%
\bibitem{YS_AdvFuncMat_05}%
  \BibitemOpen
  \bibfield{author}{%
  \bibinfo {author} {\bibfnamefont{M.~S.}\ \bibnamefont{H.~Yabu}},\ }%
  \bibfield{journal}{%
  \Doi{10.1002/adfm.200400315}{\bibinfo {journal} {Advanced Functional
  Materials}}\ }%
  \textbf{\bibinfo {volume} {15}},\ \bibinfo {pages} {575} (\bibinfo {year}
  {2005})
  \bibAnnoteFile{NoStop}{YS_AdvFuncMat_05}%
\bibitem{CLH_AccChemRes_07}%
  \BibitemOpen
  \bibfield{author}{%
  \bibinfo {author} {\bibfnamefont{X.}~\bibnamefont{Chen}}, \bibinfo {author}
  {\bibfnamefont{S.}~\bibnamefont{Lehnert}}, \bibinfo {author}
  {\bibfnamefont{M.}~\bibnamefont{Hirtz}}, \bibinfo {author}
  {\bibfnamefont{N.}~\bibnamefont{Lu}}, \bibinfo {author}
  {\bibfnamefont{H.}~\bibnamefont{Fuchs}},\ and\ \bibinfo {author}
  {\bibfnamefont{L.~F.}\ \bibnamefont{Chi}},\ }%
  \bibfield{journal}{%
  \bibinfo {journal} {Acc. Chem. Res.}\ }%
  \textbf{\bibinfo {volume} {40}},\ \bibinfo {pages} {393} (\bibinfo {year}
  {2007})%
  \bibAnnoteFile{NoStop}{CLH_AccChemRes_07}%
\bibitem{KCR_EPS_94}%
  \BibitemOpen
  \bibfield{author}{%
  \bibinfo {author} {\bibfnamefont{K.}~\bibnamefont{Spratte}}, \bibinfo
  {author} {\bibfnamefont{L.~F.}\ \bibnamefont{Chi}},\ and\ \bibinfo {author}
  {\bibfnamefont{H.}~\bibnamefont{Riegler}},\ }%
  \bibfield{journal}{%
  \bibinfo {journal} {Europhysics Letters}\ }%
  \textbf{\bibinfo {volume} {25}},\ \bibinfo {pages} {211} (\bibinfo {year}
  {1994})%
  \bibAnnoteFile{NoStop}{KCR_EPS_94}%
\bibitem{LBM_CollSurfA_00}%
  \BibitemOpen
  \bibfield{author}{%
  \bibinfo {author} {\bibfnamefont{S.}~\bibnamefont{Leporatti}}, \bibinfo
  {author} {\bibfnamefont{G.}~\bibnamefont{Brezesinski}},\ and\ \bibinfo
  {author} {\bibfnamefont{H.}~\bibnamefont{M\"ohwald}},\ }%
  \bibfield{journal}{%
  \Doi{DOI: 10.1016/S0927-7757(99)00334-9}{\bibinfo {journal} {Colloids and
  Surfaces A: Physicochemical and Engineering Aspects}}\ }%
  \textbf{\bibinfo {volume} {161}},\ \bibinfo {pages} {159 } (\bibinfo {year}
  {2000})%
  \bibAnnoteFile{NoStop}{LBM_CollSurfA_00}%
\bibitem{GR_CollSurfA_98}%
  \BibitemOpen
  \bibfield{author}{%
  \bibinfo {author} {\bibfnamefont{K.}~\bibnamefont{Graf}}\ and\ \bibinfo
  {author} {\bibfnamefont{H.}~\bibnamefont{Riegler}},\ }%
  \bibfield{journal}{%
  \Doi{DOI: 10.1016/S0927-7757(96)03923-4}{\bibinfo {journal} {Colloids and
  Surfaces A: Physicochemical and Engineering Aspects}}\ }%
  \textbf{\bibinfo {volume} {131}},\ \bibinfo {pages} {215} (\bibinfo {year}
  {1998})
  \bibAnnoteFile{NoStop}{GR_CollSurfA_98}%
\bibitem{SR_Langmuir_94}%
  \BibitemOpen
  \bibfield{author}{%
  \bibinfo {author} {\bibfnamefont{K.}~\bibnamefont{Spratte}}\ and\ \bibinfo
  {author} {\bibfnamefont{H.}~\bibnamefont{Riegler}},\ }%
  \bibfield{journal}{%
  \bibinfo {journal} {Langmuir}\ }%
  \textbf{\bibinfo {volume} {10}},\ \bibinfo {pages} {3161} \bibinfo {year} {1994})
  \bibAnnoteFile{NoStop}{SR_Langmuir_94}%
\bibitem{RS_ThinSolidFilms_92}%
  \BibitemOpen
  \bibfield{author}{%
  \bibinfo {author} {\bibfnamefont{H.}~\bibnamefont{Riegler}}\ and\ \bibinfo
  {author} {\bibfnamefont{K.}~\bibnamefont{Spratte}},\ }%
  \bibfield{journal}{%
  \Doi{DOI: 10.1016/0040-6090(92)90153-3}{\bibinfo {journal} {Thin Solid
  Films}}\ }%
  \textbf{\bibinfo {volume} {210-211}},\ \bibinfo {pages} {9} (\bibinfo {year}
  {1992})
  \bibAnnoteFile{NoStop}{RS_ThinSolidFilms_92}%
\bibitem{KGFC_Langmuir_10}%
  \BibitemOpen
  \bibfield{author}{%
  \bibinfo {author} {\bibfnamefont{M.~H.}\ \bibnamefont{K\"opf}}, \bibinfo
  {author} {\bibfnamefont{S.~V.}\ \bibnamefont{Gurevich}}, \bibinfo {author}
  {\bibfnamefont{R.}~\bibnamefont{Friedrich}},\ and\ \bibinfo {author}
  {\bibfnamefont{L.}~\bibnamefont{Chi}},\ }%
  \bibfield{journal}{%
  \bibinfo {journal} {Langmuir}\ }%
  \textbf{\bibinfo {volume} {26}},\ \bibinfo {pages} {10444} (\bibinfo {year} {2010})
  \bibAnnoteFile{NoStop}{KGFC_Langmuir_10}%
\bibitem{RMM_PRL_03}%
  \BibitemOpen
  \bibfield{author}{%
  \bibinfo {author} {\bibfnamefont{S.}~\bibnamefont{R\"udiger}}, \bibinfo
  {author} {\bibfnamefont{D.~G.}\ \bibnamefont{M\'iguez}}, \bibinfo {author}
  {\bibfnamefont{A.~P.}\ \bibnamefont{Mu\~nuzuri}}, \bibinfo {author}
  {\bibfnamefont{F.}~\bibnamefont{Sagu\'es}},\ and\ \bibinfo {author}
  {\bibfnamefont{J.}~\bibnamefont{Casademunt}},\ }%
  \bibfield{journal}{%
  \bibinfo {journal} {Phys. Rev. Lett.}\ }%
  \textbf{\bibinfo {volume} {90}},\ \bibinfo {pages} {128301} (\bibinfo {year} {2003})
  \bibAnnoteFile{NoStop}{RMM_PRL_03}%
\bibitem{BEI_RevModPhys_09}%
  \BibitemOpen
  \bibfield{author}{%
  \bibinfo {author} {\bibfnamefont{D.}~\bibnamefont{Bonn}}, \bibinfo {author}
  {\bibfnamefont{J.}~\bibnamefont{Eggers}}, \bibinfo {author}
  {\bibfnamefont{J.}~\bibnamefont{Indekeu}}, \bibinfo {author}
  {\bibfnamefont{J.}~\bibnamefont{Meunier}},\ and\ \bibinfo {author}
  {\bibfnamefont{E.}~\bibnamefont{Rolley}},\ }%
  \bibfield{journal}{%
  \Doi{10.1103/RevModPhys.81.739}{\bibinfo {journal} {Rev. Mod. Phys.}}\ }%
  \textbf{\bibinfo {volume} {81}},\ \bibinfo {eid} {739} (\bibinfo {year}
  {2009})
  \bibAnnoteFile{NoStop}{BEI_RevModPhys_09}%
\bibitem{CM_RevModPhys_09}%
  \BibitemOpen
  \bibfield{author}{%
  \bibinfo {author} {\bibfnamefont{R.~V.}\ \bibnamefont{Craster}}\ and\
  \bibinfo {author} {\bibfnamefont{O.~K.}\ \bibnamefont{Matar}},\ }%
  \bibfield{journal}{%
  \Doi{10.1103/RevModPhys.81.1131}{\bibinfo {journal} {Rev. of Mod. Phys.}}\ }%
  \textbf{\bibinfo {volume} {81}},\ \bibinfo {eid} {1131} (\bibinfo {year}
  {2009})
\bibitem{ODB_RevModPhys_97}%
  \BibitemOpen
  \bibfield{author}{%
  \bibinfo {author} {\bibfnamefont{A.}~\bibnamefont{Oron}}, \bibinfo {author}
  {\bibfnamefont{S.~H.}\ \bibnamefont{Davis}},\ and\ \bibinfo {author}
  {\bibfnamefont{S.~G.}\ \bibnamefont{Bankoff}},\ }%
  \bibfield{journal}{%
  \Doi{10.1103/RevModPhys.69.931}{\bibinfo {journal} {Rev. Mod. Phys.}}\ }%
  \textbf{\bibinfo {volume} {69}},\ \bibinfo {pages} {931} (\bibinfo {year}
  {1997})%
  \bibAnnoteFile{NoStop}{ODB_RevModPhys_97}%
\bibitem{MT_PhysFluids_97}%
  \BibitemOpen
  \bibfield{author}{%
  \bibinfo {author} {\bibfnamefont{O.~K.}\ \bibnamefont{Matar}}\ and\ \bibinfo
  {author} {\bibfnamefont{S.~M.}\ \bibnamefont{Troian}},\ }%
  \bibfield{journal}{%
  \Doi{10.1063/1.869502}{\bibinfo {journal} {Phys. Fluids}}\ }%
  \textbf{\bibinfo {volume} {9}},\ \bibinfo {pages} {3645} (\bibinfo {year}
  {1997})%
  \bibAnnoteFile{NoStop}{MT_PhysFluids_97}%
\bibitem{WCM_PhysFluids_02}%
  \BibitemOpen
  \bibfield{author}{%
  \bibinfo {author} {\bibfnamefont{M.~R.~E.}\ \bibnamefont{Warner}}, \bibinfo
  {author} {\bibfnamefont{R.~V.}\ \bibnamefont{Craster}},\ and\ \bibinfo
  {author} {\bibfnamefont{O.~K.}\ \bibnamefont{Matar}},\ }%
  \bibfield{journal}{%
  \Doi{10.1063/1.1511734}{\bibinfo {journal} {Phys. Fluids}}\ }%
  \textbf{\bibinfo {volume} {14}},\ \bibinfo {pages} {4040} (\bibinfo {year}
  {2002})%
  \bibAnnoteFile{NoStop}{WCM_PhysFluids_02}%
\bibitem{TK_PRL_06}%
  \BibitemOpen
  \bibfield{author}{%
  \bibinfo {author} {\bibfnamefont{U.}~\bibnamefont{Thiele}}\ and\ \bibinfo
  {author} {\bibfnamefont{E.}~\bibnamefont{Knobloch}},\ }%
  \bibfield{journal}{%
  \bibinfo {journal} {Phys. Rev. Lett.}\ }%
  \textbf{\bibinfo {volume} {97}},\ \bibinfo {pages} {204501} (\bibinfo {year} {2006})
  \bibAnnoteFile{NoStop}{TK_PRL_06}%
\bibitem{TBBB_EPJE_03}%
  \BibitemOpen
  \bibfield{author}{%
  \bibinfo {author} {\bibfnamefont{U.}~\bibnamefont{Thiele}}, \bibinfo {author}
  {\bibfnamefont{L.}~\bibnamefont{Brusch}}, \bibinfo {author}
  {\bibfnamefont{M.}~\bibnamefont{Bestehorn}},\ and\ \bibinfo {author}
  {\bibfnamefont{M.}~\bibnamefont{Bär}},\ }%
  \bibfield{journal}{%
  \bibinfo {journal} {The European Physical Journal E: Soft Matter and
  Biological Physics}\ }%
  \textbf{\bibinfo {volume} {11}},\ \bibinfo {pages} {255} (\bibinfo {year} {2003})
  \bibAnnoteFile{NoStop}{TBBB_EPJE_03}%
\bibitem{KGF_EPL_09}%
  \BibitemOpen
  \bibfield{author}{%
  \bibinfo {author} {\bibfnamefont{M.~H.}\ \bibnamefont{K\"opf}}, \bibinfo
  {author} {\bibfnamefont{S.~V.}\ \bibnamefont{Gurevich}},\ and\ \bibinfo
  {author} {\bibfnamefont{R.}~\bibnamefont{Friedrich}},\ }%
  \bibfield{journal}{%
  \bibinfo {journal} {EPL (Europhysics Letters)}\ }%
  \textbf{\bibinfo {volume} {86}},\ \bibinfo {pages} {66003} (\bibinfo {year}
  {2009})
  \bibAnnoteFile{NoStop}{KGF_EPL_09}%
\bibitem{PRKsync}%
  \BibitemOpen
  \bibfield{author}{%
  \bibinfo {author} {\bibfnamefont{J.~K.}\ \bibnamefont{Arkady~Pikovsky},
  \bibfnamefont{Michael~Rosenblum}},\ }%
  \emph{\bibinfo {title} {Synchronization - A universal concept in nonlinear
  sciences}}\ (\bibinfo {publisher} {Cambridge University Press},\ \bibinfo
  {year} {2003})%
  \bibAnnoteFile{NoStop}{PRKsync}%
\bibitem{DBB_JPhysCondMat_90}%
  \BibitemOpen
  \bibfield{author}{%
  \bibinfo {author} {\bibfnamefont{J}~\bibnamefont{Daillant}}, \bibinfo
  {author} {\bibfnamefont{J.~J.}\ \bibnamefont{Benattar}},\ and\ \bibinfo
  {author} {\bibfnamefont{L.}~\bibnamefont{Bosio}},\ }%
  \bibfield{journal}{%
  \bibinfo {journal} {Journal of Physics: Condensed Matter}\ }%
  \textbf{\bibinfo {volume} {2}},\ \bibinfo {pages} {SA405--SA410} (\bibinfo {year}
  {1990})
  \bibAnnoteFile{NoStop}{DBB_JPhysCondMat_90}%
\bibitem{Adamson}%
  \BibitemOpen
  \bibfield{author}{%
  \bibinfo {author} {\bibfnamefont{A.~W.}\ \bibnamefont{Adamson}},\ }%
  \emph{\bibinfo {title} {Physical Chemistry of Surfaces}}\ (\bibinfo
  {publisher} {Wiley Interscience},\ \bibinfo {year} {1990})%
  \bibAnnoteFile{NoStop}{Adamson}%
\bibitem{Pi_PRE_04}%
  \BibitemOpen
  \bibfield{author}{%
  \bibinfo {author} {\bibfnamefont{L.~M.}\ \bibnamefont{Pismen}},\ }%
  \bibfield{journal}{%
  \Doi{10.1103/PhysRevE.70.021601}{\bibinfo {journal} {Phys. Rev. E}}\ }%
  \textbf{\bibinfo {volume} {70}},\ \bibinfo {pages} {021601} (\bibinfo {month}
  {Aug}\ \bibinfo {year} {2004})%
  \bibAnnoteFile{NoStop}{Pi_PRE_04}%
\bibitem{TVA_JPhysCM_09}%
  \BibitemOpen
  \bibfield{author}{%
  \bibinfo {author} {\bibfnamefont{U.}~\bibnamefont{Thiele}}, \bibinfo {author}
  {\bibfnamefont{I.}~\bibnamefont{Vancea}}, \bibinfo {author}
  {\bibfnamefont{A.~J.}\ \bibnamefont{Archer}}, \bibinfo {author}
  {\bibfnamefont{M.~J.}\ \bibnamefont{Robbins}}, \bibinfo {author}
  {\bibfnamefont{L.}~\bibnamefont{Frastia}}, \bibinfo {author}
  {\bibfnamefont{A.}~\bibnamefont{Stannard}}, \bibinfo {author}
  {\bibfnamefont{E.}~\bibnamefont{Pauliac-Vaujour}}, \bibinfo {author}
  {\bibfnamefont{C.~P.}\ \bibnamefont{Martin}}, \bibinfo {author}
  {\bibfnamefont{M.~O.}\ \bibnamefont{Blunt}},\ and\ \bibinfo {author}
  {\bibfnamefont{P.~J.}\ \bibnamefont{Moriarty}},\ }%
  \bibfield{journal}{%
  \bibinfo {journal} {Journal of Physics: Condensed Matter}\ }%
  \textbf{\bibinfo {volume} {21}},\ \bibinfo {pages} {264016 (13pp)} (\bibinfo
  {year} {2009})
  \bibAnnoteFile{NoStop}{TVA_JPhysCM_09}%
\bibitem{CH_JChemPhys_58}%
  \BibitemOpen
  \bibfield{author}{%
  \bibinfo {author} {\bibfnamefont{J.~W.}\ \bibnamefont{Cahn}}\ and\ \bibinfo
  {author} {\bibfnamefont{J.~E.}\ \bibnamefont{Hilliard}},\ }%
  \bibfield{journal}{%
  \Doi{10.1063/1.1744102}{\bibinfo {journal} {J. Chem. Phys.}}\ }%
  \textbf{\bibinfo {volume} {28}},\ \bibinfo {pages} {258} (\bibinfo {year}
  {1958})%
  \bibAnnoteFile{NoStop}{CH_JChemPhys_58}%
\bibitem{DS_AdvChemPhys_82}%
  \BibitemOpen
  \bibfield{author}{%
  \bibinfo {author} {\bibfnamefont{H.~T.}\ \bibnamefont{Davis}}\ and\ \bibinfo
  {author} {\bibfnamefont{L.~E.}\ \bibnamefont{Scriven}},\ }%
  \bibfield{journal}{%
  \bibinfo {journal} {Adv. Chem. Phys.}\ }%
  \textbf{\bibinfo {volume} {49}},\ \bibinfo {pages} {357} (\bibinfo {year}
  {1982})%
  \bibAnnoteFile{NoStop}{DS_AdvChemPhys_82}%
\bibitem{Ev_AdvPhysics_79}%
  \BibitemOpen
  \bibfield{author}{%
  \bibinfo {author} {\bibfnamefont{R.}~\bibnamefont{Evans}},\ }%
  \bibfield{journal}{%
  \bibinfo {journal} {Adv. Phys.}\ }%
  \textbf{\bibinfo {volume} {28}},\ \bibinfo {pages} {143} (\bibinfo {year}
  {1979})%
  \bibAnnoteFile{NoStop}{Ev_AdvPhysics_79}%
\bibitem{NRC}%
  \BibitemOpen
  \bibfield{author}{%
  \bibinfo {author} {\bibfnamefont{W.~H.}\ \bibnamefont{Press}},\ }%
  \emph{\bibinfo {title} {Numerical Recipes in C}}\ (\bibinfo {publisher}
  {University Press},\ \bibinfo {address} {Cambridge},\ \bibinfo {year}
  {1999})%
  \bibAnnoteFile{NoStop}{NRC}%
\end{thebibliography}
\end{document}